%% file: main.tex
\documentclass[reprint,superscriptaddress,amsmath,amssymb,aps,prl,colorlinks]{revtex4-2}
\usepackage{graphicx}
\usepackage{dcolumn}
\usepackage{bm}
\usepackage[usenames,dvipsnames]{color}
\usepackage{natbib}
\usepackage{amsmath}
\usepackage{amssymb}
\usepackage{booktabs}
\usepackage{textcomp}
\usepackage{soul}
\usepackage{empheq}
\usepackage[urlcolor=NavyBlue,citecolor=NavyBlue,linkcolor=NavyBlue]{hyperref}

\usepackage[normalem]{ulem}

\usepackage{verbatim}

\newcommand{\detailtexcount}[1]{%
  \immediate\write18{texcount -merge -q #1.tex > #1.wcdetail}%
  \verbatiminput{#1.wcdetail}%
}

\renewcommand{\vec}[1]{\mathbf{ #1 }}

\newcommand{\vect}[1]{\mathbf{ #1 }}

\makeatletter

\begin{document}

\title{Multi-scale patterns arising from competing chemotactic interactions}

\author{Agniva Datta}
\affiliation{Institute of Physics and Astronomy, University of Potsdam, D-14476 Potsdam, Germany}

\author{Robert Gro{\ss}mann}
\affiliation{Institute of Physics and Astronomy, University of Potsdam, D-14476 Potsdam, Germany}

\author{Carsten Beta}
\email[Correspondence should be addressed to ]{carsten.beta@uni-potsdam.de}
\affiliation{Institute of Physics and Astronomy, University of Potsdam, D-14476 Potsdam, Germany}
\affiliation{Nano Life Science Institute (WPI-NanoLSI), Kanazawa University, Kakuma-machi, Kanazawa 920-1192, Japan}

\begin{abstract}
Classical chemotaxis models consider the directed motion of cells in response to a chemical gradient.
Recent experiments suggest that biological microswimmers such as bacteria can perform chemotaxis toward multiple chemicals, which dynamically evolve depending on the distribution of the microswimmers.
In this Letter, we consider a mass-conserving chemotaxis model in which pattern formation is driven by two competing chemicals, autoinducers and nutrients, where the former acts as an activator and the latter as an inhibitor for coarsening. 
We show that this can give rise to diverse patterns including long wavelength uninterrupted coarsening, Turing-like arrested coarsening, traveling waves as well as a novel class of mixed multi-scale oscillatory coarsening patterns, where a dense aggregate coexist with a dynamically spreading part of the population. 
We analytically derive the conditions for the onset of such patterns and discuss their physical implications, which significantly broadens our understanding of signaling active matter and pattern formation in general.
\end{abstract}

\date{\today}
\pacs{}
\maketitle

A key property of many microorganisms is their ability to perform chemotaxis, the directed movement in response to chemical gradients.
Over the past decades, chemotaxis has been studied extensively for bacteria~\cite{adler1975chemotaxis,berg1975chemotaxis} as well as for amoeboid cells~\cite{van2004chemotaxis,swaney2010eukaryotic} such as \textit{Dictyostelium discoideum}, neutrophils, or macrophages~\cite{kolaczkowska2013neutrophil}.
Single-cell trajectory experiments have revealed that bacteria, moving in a run-and-turn fashion, perform chemotaxis by biasing their run durations or turning angles to follow chemical gradients, both in bulk fluid~\cite{berg1972chemotaxis,macnab1972,Alirezaeizanjani2020,Pohl2017,Seyrich2018,gomeznava2020novel,seyrich2019traveling,Nakai2021} as well as in porous media~\cite{Bhattacharjee2021,beier2026dual}.
On a population level, gradients that dynamically arise from the consumption of nutrients may result in collective migration of bacteria leading to the formation of traveling chemotactic pulses~\cite{adler1966chemotaxis,grigson2026co}. 
Moreover, many bacteria show chemotactic responses to signaling substances that they release themselves (autochemotaxis).
This often leads to aggregation and eventually biofilm formation, which has been reported for multiple bacterial species including \textit{Escherichia coli}~\cite{budrene1991complex,budrene1995dynamics} and \textit{Salmonella} Typhimurium~\cite{woodward1995spatio,blat1995tar}.
A similar phenomenon is observed in amoeboid cells such as \textit{Dictyostelium discoideum}, which show migration toward self-secreted cyclic AMP during starvation to form multicellular aggregates~\cite{ fisher1989quantitative,song2006dictyostelium,nichols2015chemotaxis}. 
Taken together, two competing mechanisms may arise:~while self-generated gradients due to nutrient consumption result in population spreading and the collective formation of traveling pulses (dispersal), the production of autochemoattractants~(autoinducers) promotes an attractive feedback that leads to coarsening patterns~(aggregation). 
This is supported by recent experiments indicating that bacteria such as \textit{E.~coli} can indeed chemotactically respond to both autoinducers and nutrients~\cite{park2003influence,scheidweiler2024spatial, phan2024direct}.

From a theoretical point of view, the most common quantitative descriptions of chemically guided motility phenomena are provided by different variants of the Keller-Segel model.
The initial model was proposed to describe chemotactic aggregation of slime moulds~\cite{keller1970initiation}, but it was subsequently generalized and modified to explain chemotaxis across a wide range of motile unicellular organisms~\cite{keller1971model}, including traveling bands in chemotactic bacteria~\cite{keller1971traveling}. 
Also Keller-Segel variants incorporating the impact of two distinct chemicals were proposed, but here the chemicals do not compete, or at most one of them acts as a chemoattractant~\cite{brenner1998physical,tyson1999minimal,ma2014dynamic,cremer2019chemotaxis}.
Related work has addressed phase separation of chemotactic or chemokinetic active particles~\cite{zhao2023chemotactic,kwon2025phase,rasshofer2025capillary}, non-reciprocal pattern formation in multi-species mixtures mediated by quorum sensing and chemotaxis~\cite{grauer2020swarm,dinelli2023non,zhao2026dynamic,liu2024self}, as well as cell-cell communication via autoinducers that was shown to modify bacterial chemotaxis along an externally imposed gradient~\cite{long2017cell,saha2026effects}.
The effect of competing gradients has been considered only in one-dimensional studies of traveling and solitary pulses~\cite{salman2006solitary,saragosti2010mathematical,young2019interactions} and in particle-level models of chemotaxis in neutrophils~\cite{perez2026competing} and bacteria~\cite{curk2013chemotactic}.
How competing gradients of a self-generated autochemotactic signal and a locally consumed nutrient field jointly shape the complex population patterns of active particles has not been systematically explored to date.

\begin{figure*}[t]
 	\includegraphics[width=\textwidth]{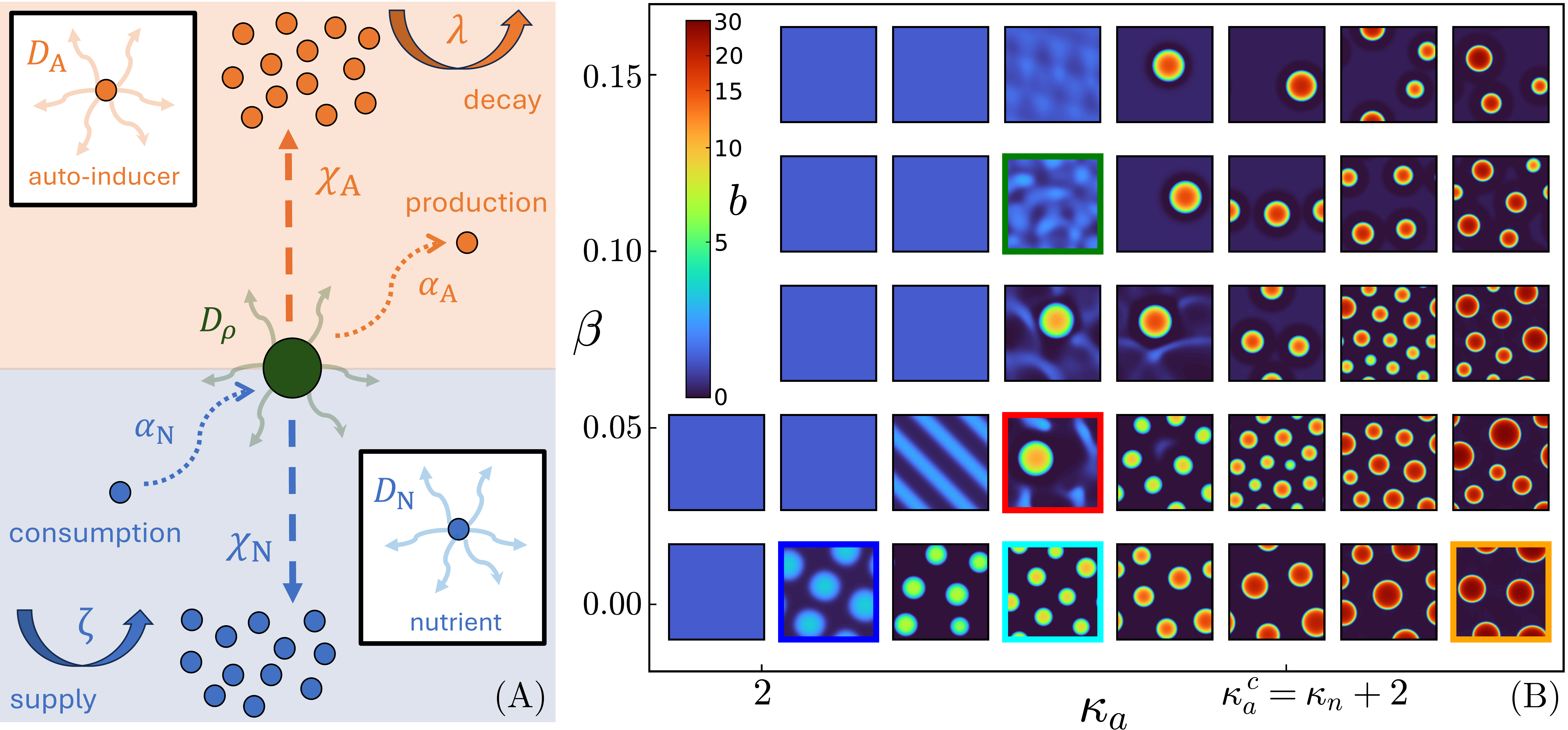}
	\vspace*{-0.55cm}
	\caption{Illustration of the model dynamics and emerging patterns. (A)~Schematic of the model dynamics. (B) Overview of emerging patterns. Numerical simulations of the model~[cf.~Eq.~\eqref{eq:model}] with periodic boundary conditions starting from the homogeneous steady state with small fluctuations reveal a wide range of patterns, including traveling waves, hexagonal arrays of spots, various types of coarsening as well as coexistence of the traveling spots and waves. The control parameters set the response of particles to autoinducers~($\kappa_a$) and the inverse timescale of the nutrient dynamics~($\beta$). Simulation parameters for the non-dimensionalized model Eq.~\eqref{eq:model}:~$\kappa_n =14$,~$\delta_n=0.75$,~$\delta_a=2$~(see main text for definition), system size~$L = 50$. See also Supplementary Video SV1~\cite{noteSI}. } 
  	\label{fig:kinetics_all_patterns}
 \end{figure*}

\paragraph{Model}

In this article, we consider a mass-conserved three-component Keller-Segel model, in which the motility of cells is chemotactically coupled to two dynamically distinct competing chemicals, a self-generated autochemotactic species (autoinducer) and a locally consumed nutrient field.
The model is illustrated schematically in Fig.~\ref{fig:kinetics_all_patterns}A. 
We assume that the density of particles~$\rho(\vect{x},t)$ evolves according to a Fokker-Planck equation including drift and diffusion,~$\partial_t\rho\left(\vect{x},t\right) = D_\rho \nabla^2 \rho - \nabla \cdot \vect{J}$, in which~$D_\rho$ is the effective diffusivity of cells and~$\vect{J}= \vec{J}_A + \vec{J}_N$ is a chemotactic flux toward autoinducers~($C_A$) and nutrients~($C_N$)~\cite{saha2026effects}. 
The individual terms read~$\vect{J}_{A,N} = \rho \!\,\, \tilde{\chi}_{A,N}\left(\rho,C_{A,N}\right) \nabla C_{A,N}$. 
The specific structure of the chemotactic sensitivity~$\tilde{\chi}_{A,N}\left(\rho,C_{A,N}\right)$ depends on the model variant considered. 
Here, we employ the volume-filling approximation~$\tilde{\chi}_{A,N}=\chi_{A,N}\rho_{\rm th}^2/\left(\rho^2+\rho_{\text{th}}^2\right)$ via a second-order hill function in~$\rho$ to prevent overcrowding of cells that often leads to chemotactic collapse~\cite{hillen2001global,painter2002volume,childress1984chemotactic,herrero1996chemotactic}.
The threshold density~$\rho_{\rm th}$ is set to be equal to the spatially homogeneous cell density~$\rho_0$; the parameters~$\chi_{A,N}$ control the chemotactic response. 
For simplicity, we do not consider cell proliferation, so that the particle number is conserved.

Autoinducers~$C_A$~(diffusion coefficient~$D_A$) are produced by the cells at a rate~$\alpha_A$ and decay at a rate~$\lambda$, leading to the dynamics~\cite{childress1981nonlinear,saha2026effects}:~$\partial_tC_A\left(\vect{x},t\right) = D_A \nabla^2 C_A + \alpha_A \rho - \lambda C_A$. 
On the other hand, nutrients~$C_N$~(diffusivity~$D_N$) are consumed by the cells at a rate~$\alpha_N$ and supplied externally at a constant rate $\zeta$ uniformly in space~\cite{zhao2023chemotactic}:~$\partial_tC_N\left(\vect{x},t\right) = D_N \nabla^2 C_N - \alpha_N \rho \!\; C_N + \zeta$. 
After non-dimensionalizing with respect to the homogeneous steady states~$\left(b,a,n\right)=\left(\rho/\rho_0, \ C_A/C_A^0, \ C_N/C_N^0\right)$, and rescaling time and length by~$\tau = 1/\lambda$ and~$\ell=\sqrt{D_\rho/\lambda}$, we arrive at the following reduced model, 
\begin{subequations}
\label{eq:model}
\begin{align}
	\!\!    \partial_tb\left(\vect{x},t\right) &=  \nabla^2 b -\vect{\nabla} \! \cdot \! \left[\frac{\kappa_{a} b}{1+b^2} \nabla a\right] \! -\vect{\nabla} \! \cdot \! \left[\frac{\kappa_{n} b}{1+b^2} \nabla n\right] \!\!\, , \!\! \label{eqn:dynamics_b}\\
	\!\!    \partial_ta\left(\vect{x},t\right) &=  \delta_a \nabla^2 a + b - a, \label{eqn:dynamics_a}\\
	\!\!    \partial_tn\left(\vect{x},t\right) &=  \delta_n \nabla^2 n - \beta bn + \beta, \label{eqn:dynamics_n}
\end{align}
\end{subequations}
where the relevant system parameters are the effective chemotactic coefficients~$\kappa_{a,n} = \chi_{A,N} C_{A,N}^0/{D_\rho}$ and relative diffusivities~$\delta_{a,n} = D_{A,N}/{D_\rho}$ of the autoinducer and the nutrient, respectively, and the effective~(inverse) timescale of the nutrient dynamics~$\beta= \alpha_N \rho_0/\lambda$.
We choose~$\beta$ and the coupling strength~$\kappa_a$, in relation to~$\kappa_n$, as the relevant control parameters for exploring the phenomenology of the system.

\paragraph{Phenomenology}

We first report the results of numerical simulations in the~$\left(\kappa_a-\beta\right)$ parameter plane, keeping~$\kappa_n$, $\delta_a$, and $\delta_n$ fixed, with periodic boundary conditions in two dimensions, revealing a diverse variety of patterns summarized in Fig.~\ref{fig:kinetics_all_patterns}B. 
Here, we focus on the case of fast autoinducer diffusion~($\delta_a>\delta_n$) leading to stronger gradients in nutrients.
This potentially favors the formation of nontrivial patterns beyond classical coarsening known from the Keller-Segel model, due to the complex interplay of chemotaxis toward both, autoinducers and nutrients.

Classical uninterrupted coarsening is observed if the coupling to autoinducers dominates~($\kappa_a \gg \kappa_n$); cf.~rightmost panels, e.g., orange inset in Fig.~\ref{fig:kinetics_all_patterns}B. 
In contrast, a plethora of novel patterns arises if both couplings are comparable~(more specifically for parameters $\kappa_a \lesssim \kappa_n$). 
For small values of~$\beta$~(fast nutrient dynamics), the nutrient concentration is small in areas of high cell density, and chemotaxis to nutrients implies a flux of particles away from these regions.
This mechanism arrests coarsening so that, in consequence, multiple aggregates get locally and eventually globally ordered in hexagonal lattices, see bottom row, e.g., cyan and blue insets of Fig.~\ref{fig:kinetics_all_patterns}B. 
This is reminiscent of the observation that chemotaxis toward nutrients may prevent aggregation of active Brownian particles~\cite{zhao2023chemotactic}.
\begin{figure}[t]
    \centering
    \includegraphics[width=\linewidth]{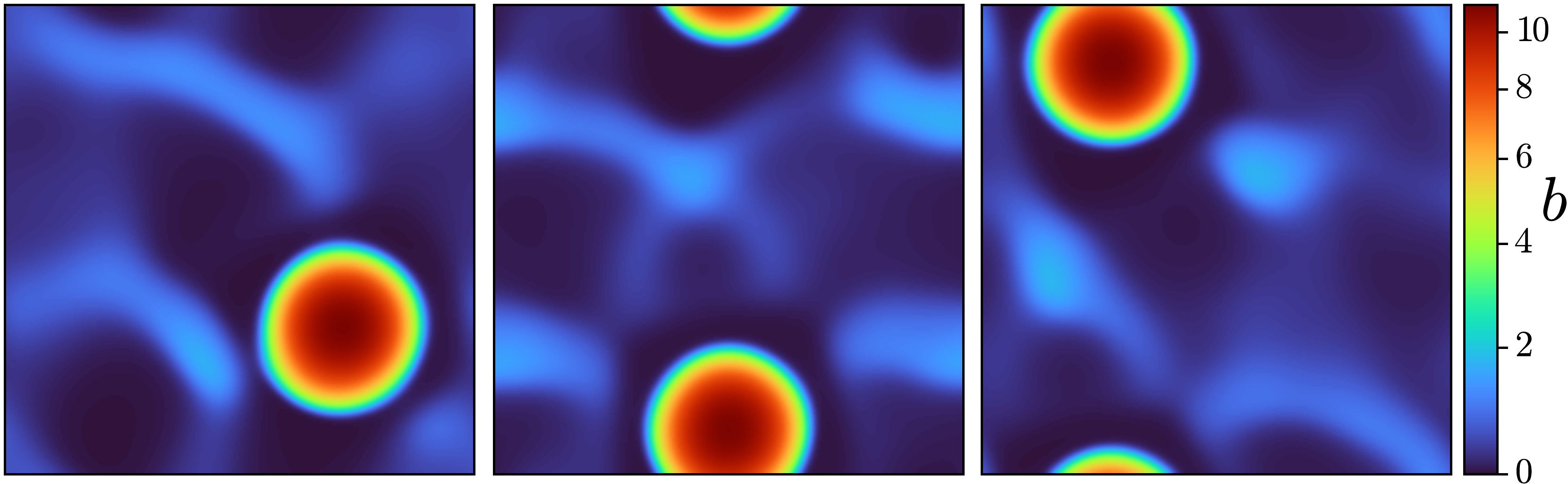}
    \vspace*{-0.55cm}
    \caption{Multi-scale pattern showing an exceptional instability giving rise to coexistence: A locally formed high-density droplet of particles moves in a background of traveling bands of cells. Time increases from left to right. Simulation parameters: ~$\kappa_a=5$, $\kappa_n=14$, $\delta_a=2.5$, $\delta_n=0.75$, $\beta=0.02$, system size~$L = 50$. See also Supplementary Video SV2~\cite{noteSI}.} 
    \label{fig:ballistic_coarsening}
\end{figure}
\begin{figure*}[t]
 	\includegraphics[width=\textwidth]{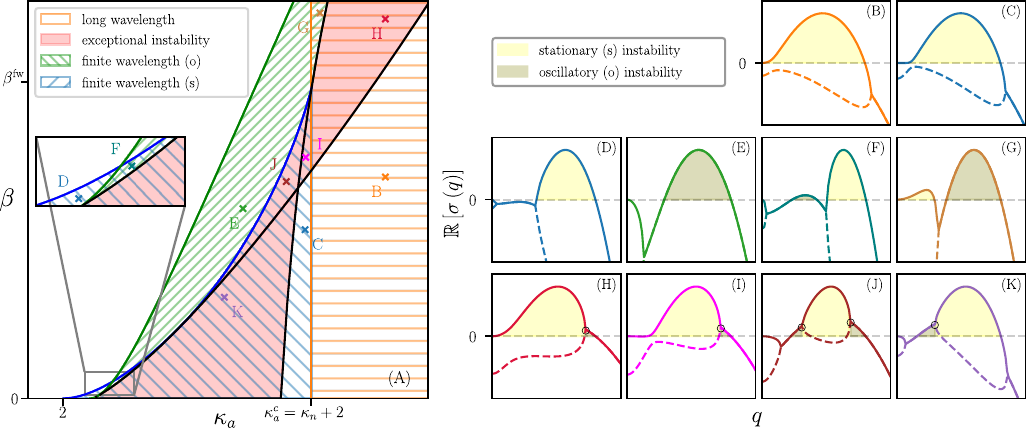}
	\vspace*{-0.55cm}
	\caption{Phase diagram and dispersion relations. (A) The~$\left(\kappa_a-\beta\right)$ phase diagram is shown for~$\kappa_n =14$, $\delta_n=0.75$, $\delta_a=2.5$. The white color correspond to regions where the system is linearly stable. (B-K) Dispersion curves corresponding to the points B-K in panel~(A). The solid lines correspond to the leading~(real part of) eigencurves and the dashed lines correspond to the second eigencurves plotted; they are plotted in their respective colors from panel~(A). Stationary instabilities~(yellow shading) arise when the leading eigencurve remains real and positive, while oscillatory instabilities~(olive shading) arise for pairs of complex conjugate eigenvalues. The third eigencurves monotonically decrease without ever becoming positive and hence not shown. In panels~(B-D), the dispersion relations are associated with stationary instabilities: a long wavelength instability and two instances~(in different parameter regions) of finite wavelength instability, respectively. Panel~(E) shows the dispersion relation for an oscillatory finite wavelength instability. Panels~(F, G) show the coexistence of stationary and oscillatory instabilities occurring independently. The dispersion curves in panels~(H-K) correspond to the different points in the red shaded region in panel~(A), highlighting exceptional instabilities: at least one of the exceptional points (highlighted by black circles) lies above the~$\mathbb{R}[\sigma]=0$-axis, thereby giving rise to a mixed band of unstable oscillatory and stationary modes.}  
     \label{fig:phase_space_dispersion}
 \end{figure*}
As~$\beta$ increases, the nutrient dynamics starts to dominate, and if coupling to autoinducers~($\kappa_a$) is small compared to the response to nutrients~($\kappa_n$), the formation of aggregates is suppressed and traveling bands arise instead~(green inset in Fig.~\ref{fig:kinetics_all_patterns}B). 
However, for slightly larger~$\kappa_a$, complex mixed patterns are observed.
Here, a stable aggregate forms within a background of traveling waves, which remains stationary or gets carried along with the traveling bands, either ballistically or in a random fashion (see red inset in Fig.~\ref{fig:kinetics_all_patterns}B). 
We specifically highlight snapshots of a time evolution of this novel composite pattern in Fig.~\ref{fig:ballistic_coarsening}.

\paragraph{Linear stability analysis}

We analytically characterize the conditions for the onset of the various types of patterns by linear stability analysis. 
To this end, the model dynamics~[Eq.~\eqref{eq:model}] for the system state~$\vect{B} \equiv \left(b,a,n\right)$ was linearized around the homogeneous steady state~$\vect{B_0} \equiv \left(1,1,1\right)$ and expressed in terms of Fourier modes~(wave-number~$q$). 
The dispersion relation~$\sigma(q)$ can be obtained from
\begin{align}
\label{eq:disp_rel}
&\left(\sigma + q^2\right)\left(\sigma + 1 + \delta_a q^2\right)\left(\sigma + \beta + \delta_n q^2\right) - \\
& \ \ \ \ \frac{\kappa_a}{2} q^2\left(\sigma + \beta + \delta_n q^2\right) + \frac{\kappa_n}{2} q^2 \beta\left(\sigma + 1 + \delta_a q^2\right) = 0, \nonumber
\end{align}
for details, see Supplemental Material~\cite{noteSI}.
Here,~$\sigma(q)$ correspond to the eigenvalues of the linear system determining the stability.
Whenever the real part of the eigencurves~$\sigma(q)$ become positive~($\mathbb{R} \left[ \sigma(q) \right] > 0$), an instability is expected. 
In order to identify the bands of unstable modes, we set~$\mathbb{R}\left[\sigma\! \left(q\right)\right] = 0$ in the dispersion relation and look for real and positive values of~$q$, which satisfy the dispersion relation.
This allows us to analytically derive the conditions for stationary~(purely real~$\sigma(q)$) and oscillatory instabilities~(nonzero imaginary part of $\sigma(q)$) and express them as boundaries in the~$\left(\kappa_a-\beta\right)$ parameter space, see Supplemental Material for derivations~\cite{noteSI}. 
The analytically derived boundaries, the instability regions, and exemplary dispersion relations are summarized in Fig.~\ref{fig:phase_space_dispersion}.

We find two types of stationary instabilities:~a long wavelength instability (Fig.~\ref{fig:phase_space_dispersion}B) and a finite wavelength Turing-like instability (Fig.~\ref{fig:phase_space_dispersion}C, D).
The condition for the long wavelength instability, reminiscent of the classical Keller-Segel model, is given by~$\kappa_a - \kappa_n > 2$. 
One can equivalently derive this condition by expanding~$\sigma(q)$ for small~$q$ around~$\sigma(q)=0$, i.e.~the leading eigencurve~\cite{noteSI}.
This instability explains the coarsening patterns as observed numerically~(Fig.~\ref{fig:kinetics_all_patterns}B). 
Note that, in the absence of the nutrient dynamics, $\kappa_a>2$ is the condition for a long wavelength instability in the classical Keller-Segel model with volume-filling approximation. 
The transition to coarsening is thus shifted to larger values of~$\kappa_a$ as chemotaxis toward nutrients hinders coarsening.

Turing-like instabilities, leading to stationary finite wavelength patterns, are observable in our system only in the parameter band~$2<\kappa_a<\kappa_n + 2$.
Hence, they only arise due to the finite chemotactic drift toward nutrients~($\kappa_n>0$) as a genuine result of the competing effects of chemotaxis toward two distinct chemical species. 
Specifically, we find a parameter window of width~$\kappa_a^{\text{fw}}=\kappa_n$ and height~$\beta^{\text{fw}}=\delta_n \kappa_n/\left[\delta_a\!\left(2+\kappa_n\right)\right]$ in which these instabilities can emerge.
Accordingly, a faster nutrient diffusion and a slower autoinducer diffusion promote arrested coarsening, provided~$\kappa_n$ is sufficiently large. 
Thus, the instability mechanism resembles an activator-inhibitor system, in which autoinducers act as a \textit{coarsening activator} while the nutrients take the role of a \textit{coarsening inhibitor}.

We found two regions in parameter space in which finite wavelength instabilities are expected, highlighted in blue in Fig.~\ref{fig:phase_space_dispersion}A. 
One parameter regime, exemplified by the dispersion relation in Fig.~\ref{fig:phase_space_dispersion}C, is sandwiched between different other instability regions~(higher order bifurcations described below), and hence carry their strong nonlinear effects, which leads to spatially arranged clusters of different sizes, as observed in numerical simulations~(cf.~Fig.~\ref{fig:kinetics_all_patterns}B). 
However, the instability exemplified by Fig.~\ref{fig:phase_space_dispersion}D arises directly from the stable region as a primary bifurcation and, therefore, Turing spots emerge which are arranged in hexagonal lattices; cf.~bottom row of Fig.~\ref{fig:kinetics_all_patterns}B.

Beyond stationary instabilities, we find one purely oscillatory instability in the system, which occurs for a finite wavelength (Fig.~\ref{fig:phase_space_dispersion}E). 
The patterns in this region typically show multiple modes of traveling waves interfering with each other~(green inset in Fig.~\ref{fig:kinetics_all_patterns}B). 
In addition, various types of mixed instabilities were found.
As shown in Figs.~\ref{fig:phase_space_dispersion}F and~G, we obtain independent coexistence of a finite wavelength oscillatory instability with either a finite wavelength stationary instability or a long wavelength stationary instability, respectively.
In Fig.~\ref{fig:phase_space_dispersion}F, the fastest growing mode is that of a stationary instability but the characteristic wavelength of the oscillatory instability is larger; Fig.~\ref{fig:phase_space_dispersion}G shows the opposite case. 
The most interesting situation arises in the red-shaded region in Fig.~\ref{fig:phase_space_dispersion}A, where the stationary and oscillatory instabilities coexist within the same band.
Such mixed instabilities appear as eigencurves branch and split, giving rise to exceptional points above the~$q$-axis as shown by black circles in Fig.~\ref{fig:phase_space_dispersion}H-K.
This type of multi-scale mixed instability has been recently observed also in non-reciprocal pattern forming systems~\cite{brauns2024nonreciprocal,winkler2026simple}.
The pattern highlighted in Fig.~\ref{fig:ballistic_coarsening} is an example from this instability region, corresponding to the dispersion relation shown in Fig.~\ref{fig:phase_space_dispersion}K.
Here, the oscillatory modes of larger characteristic length scale survive in the background and drive the motion of the aggregate, which is formed as a result of nonlinear effects stemming from the conservation of cell density.

\paragraph{Coalescence vs.~ripening of aggregates}

Although the condition for uninterrupted coarsening patterns observed for~$\kappa_a>\kappa_n+2$ in Fig.~\ref{fig:kinetics_all_patterns}B can be rationalized by the linear stability analysis as a consequence of a long wavelength instability~(Fig.~\ref{fig:phase_space_dispersion}B), one can observe two different types of coarsening patterns at the nonlinear level.
While for low~$\beta$, the aggregates coalesce to form bigger aggregates, for increasing~$\beta$, the aggregates start to show ripening, i.e., larger aggregates grow bigger and smaller aggregates shrink to eventually disappear. 
Here, we quantitatively characterize these two coarsening regimes relying on the principle of a mass-competition instability and singular perturbation theory~\cite{brauns2021wavelength,weyer2023coarsening,frey2026pattern};~see also End Matter for details.
We define a mass-redistribution potential~$\eta \! \left(b,a,n\right)$ for our model, so that Eq.~(\ref{eqn:dynamics_b}) can be rewritten as a continuity equation,~$\partial_t b+\vect{\nabla} \cdot \vect{J}_{\text{eff}}=0$, with the effective flux~$\vect{J}_{\text{eff}}=-\mu  \! \left(b\right) \nabla \eta$ that drives cells toward regions of lower values of $\eta$ with an effective mobility $\mu \! \left(b\right)$ given by~$\mu \! \left(b\right)=b/ \! \left(1+b^2\right)$~\cite{weyer2025coarsening}.

\begin{figure}[t]
    \centering
    \includegraphics[width=\linewidth]{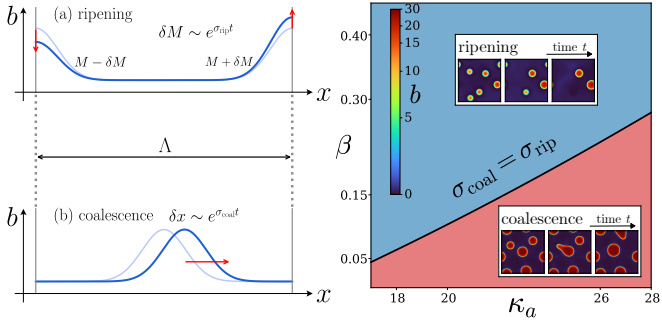}
    \vspace*{-0.6cm}
    \caption{Coalescence and ripening. Estimate of the transition line~$\sigma_{\rm coal}=\sigma_{\rm rip}$ from singular perturbation theory. While the blue shaded region~$(\sigma_{\rm coal}<\sigma_{\rm rip})$ is dominated by ripening patterns, the red shaded region~$(\sigma_{\rm coal}>\sigma_{\rm rip})$ is dominated by coalescence patterns. The color bar is associated with the density~$b$ in snapshots of simulated patterns shown for $(\kappa_a, \beta)$ values lying in the two predicted regions. See also Supplementary Video SV3~\cite{noteSI}.} 
    \label{fig:coarsening_patterns}
\end{figure}

In order to estimate the relative tendency of neighboring aggregates to coalesce or ripen, we use a local quasi-steady-state approximation where the shape of an aggregate and the associated chemical fields are assumed to relax rapidly in comparison to the coarsening dynamics of peak positions and peak masses.
Following Refs.~\cite{weyer2025coarsening,weyer2025chemotaxis}, we approximate a local pair of neighboring aggregates by a stationary half-peak profile in one dimension on the half line~$x\in[0,\Lambda/2]$, where~$(x=0)$ denotes the center of a density peak and~$x=\Lambda/2$ corresponds to the surrounding low-density plateau (see Fig.~\ref{fig:coarsening_patterns}).
The stationary profile satisfies the no-flux condition~$\partial_x\eta=0$, together with reflective boundary conditions for the chemical fields and the mass constraint~$\int_0^{\Lambda/2} b(x)\,dx=M/2$.
We solve the auxiliary boundary value problem numerically following Eqs.~(\ref{eqn:dynamics_a}, \ref{eqn:dynamics_n}) and define~$b_+=b(0)$ and~$b_-=b(\Lambda/2)$.
Similarly, in order to estimate the sensitivity of the stationary redistribution potential~$\eta_{\text{stat}}(M)$ to the peak mass, we solve the boundary value problem for the nearby stationary profiles with masses~$M\pm\Delta M$, giving~$\partial_M\eta_{\rm stat}\simeq[\eta_{\rm stat}(M+\Delta M)-\eta_{\rm stat}(M-\Delta M)]/(2\Delta M)$.
The ripening mode is then interpreted as a mass-redistribution instability between neighboring aggregates while the coalescence mode is interpreted as a displacement instability of neighboring peaks.
The corresponding growth rates are given by
\begin{subequations}
\label{eq:growth_rates}
\begin{align}
	\sigma_{\text{rip}} &=  - \dfrac{4}{\Lambda} \mu(b_{-}) \partial_M \eta_{\text{stat}}\left(M\right) \! ,\label{eq:sigma_rip}  \\
	\sigma_{\text{coal}} &=  -2 \mathcal{G}_{\text{peak}}\left(b_{+} - b_{-}\right) \partial_M \eta_{\text{stat}}\left(M\right) \! . \label{eq:sigma_coal}
\end{align}
\end{subequations}
From these growth rates, we find the transition line given by~$\sigma_{\text{rip}}=\sigma_{\text{coal}}$ in the long wavelength instability region of the~$\left(\kappa_a-\beta\right)$ space, which provides a local rate-based mechanistic estimate of the crossover between displacement-driven coalescence and mass-redistribution-driven ripening of the chemotactic aggregates~\cite{transition_line}, as shown in Fig.~\ref{fig:coarsening_patterns}B and confirmed by direct numerical simulations.

\paragraph{Discussion}

In this Letter, we studied a multi-species Keller-Segel model in which chemotaxis is governed by the competition of two dynamically evolving chemical gradient fields.
Systematic numerical simulations show that the model produces rich spatiotemporal patterns.
Linear stability analysis traces these patterns to stationary instabilities, oscillatory instabilities, and combinations of both.
Using singular perturbation theory, we furthermore derived the conditions for the onset of two kinds of uninterrupted coarsening, coalescence and ripening, and validated them numerically.

Besides the coarsening patterns known from the classical Keller-Segel model, we observe Turing-type periodic spots and traveling waves.
Moreover, we discovered novel patterns in which aggregates coexist with a background of traveling bands.
These dynamical states are of particular biological relevance, as in bacteria such states would allow aggregation to set in, while at the same time, the population keeps exploring regions far from the aggregate via dynamic density waves~\cite{mcdougald2012should,laganenka2016}.
Such patterns resemble the dynamic clustering found in several particle-level descriptions of chemotaxis~\cite{lushi2018nonlinear,theurkauff2012dynamic,pohl2014dynamic,liebchen2017phoretic,liebchen2018synthetic,stark2018artificial} and, more generally, in active phase-separating systems~\cite{agudo2019active,cates2025active,osmanovic2025complex,willems2025run,oza2026traveling}. 
Note also that bacterial aggregation plays a key role during the early stages of biofilm formation, which governs many medically relevant processes, including diseases such as cystic fibrosis~\cite{borisova2025biofilm}.
Because the emergent patterns depend on nutrient supply, this offers a promising route to controlling aggregation, readily accessible in experiments.

A key next step is the experimental observation of the predicted patterns.
Both artificial microswimmers that mimic chemotaxis~\cite{jin2017chemotaxis,nava2018markovian,peng2024self,xiong2025chemotactic,raman2026clustering,wang2026microscopic}, as well as biological microswimmers whose growth and proliferation timescales are much longer than the timescales of pattern formation in our model offer promising experimental perspectives.
Besides, our model is also relevant to therapies that rely on disrupting bacterial quorum sensing by supplying autoinducer mimics: the nutrient field can be reinterpreted as an externally supplied mimic that the bacteria consume but that does not contribute to their growth~\cite{oloughlin2013quorum}.

On the theoretical side, the nonlinear dynamics near the onset of mixed stationary-oscillatory instabilities, arising from secondary and higher-order bifurcations, can be studied using bifurcation theory and numerical continuation.
In addition, to describe cell dynamics with growth and proliferation timescales comparable to those of our system, the model can be extended with logistic terms~\cite{cates2010arrested,centler2011modeling,gelimson2015collective,mukherjee2018growth,weyer2025chemotaxis}. 
Including growth would connect our model to classic works on pattern formation in expanding bacterial cultures~\cite{budrene1991complex,budrene1995dynamics,woodward1995spatio,blat1995tar} and may give further insight into the underlying mechanisms.
In general, proliferation terms will break the conservation law in our model and are expected to yield an even richer variety of nonlinear patterns, such as spiral waves, target patterns, and breathing spots (see Supplementary Video SV4~\cite{noteSI} for examples), which we leave for future studies.

\begin{acknowledgments}

\paragraph{Acknowledgments}

This research has been partially funded by Deutsche Forschungsgemeinschaft~(DFG): Project-ID 318763901--SFB1294. 

\end{acknowledgments}

\input{main.bbl}

\section*{End Matter}

\paragraph{Mass redistribution potential}

The properties and physical interpretation of the mass redistribution potential~$\eta \! \left(b,a,n\right)$ are discussed in this section. Comparing~$\partial_t b=\vect{\nabla} \cdot \left[\mu\!\left(b\right)\! \nabla \eta\right]$ with Eq.~(\ref{eqn:dynamics_b}), we find 
\begin{align}
	\eta\left(b,a,n\right)= \ln{b} + \dfrac{b^2}{2} - \kappa_a a - \kappa_n n.  \label{eq:redistribution_potential}    
\end{align}
The value of~$\eta$ increases with cell density~$b$ that drives diffusion of cells, while it decreases with the chemical densities~$a$ and~$n$ that drives chemotactic aggregation. Due to production and consumption of the two chemicals, perturbations in the local cell density~$b$ affect the chemical densities~$a$ and~$n$ as par Eq.~(\ref{eqn:dynamics_a}) and Eq.~(\ref{eqn:dynamics_n}), respectively. So, the condition for aggregation dominating over diffusion of cells, i.e., uninterrupted coarsening, will require a decrease in~$\eta$ with an effective increase in cell density~$b$. For long-wavelength modulations, the mass redistribution is slow compared to the local relaxation dynamics of the two chemicals~$a$ and~$n$ which rapidly adjusts to the quasi-steady state~$(a,n) \sim (b,1/b)$, according to Eq.~(\ref{eqn:dynamics_a}) and Eq.~(\ref{eqn:dynamics_n}), defined by the nullcline~$\eta^{*}\left(b\right)=\eta \left(b,b,1/b\right)$. Therefore, starting from the homogeneous steady state (hss) given by~$(b,a,n) = (1,1,1)$, the condition for uninterrupted coarsening is given by~$\partial_b \eta^{*}\left(b_{\text{hss}}\right)<0$, which is the nullcline-slope criterion~\cite{weyer2025chemotaxis}. From Eq.~(\ref{eq:redistribution_potential}), one can directly see that the nullcline-slope criterion leads to the same criterion for long wavelength instability as derived independently from the dispersion relation obtained by linear stability analysis. 

\paragraph{Ripening and coalescence growth rates} 
In one dimension, the effective flux~$\vect{J}_{\text{eff}}$ expressed as~$J_{\text{eff}}=-\mu \! \left(b\right) \! \partial_x \eta$, can be interpreted as a conductance law---gradients of the potential~$\eta$ drive cell-density fluxes, and the mobility~$\mu \!\left(b\right)$ acts as an effective conductance density. For a quasi-stationary current through a region~$\Omega$, the flux can be related directly to the potential difference~$\Delta \eta$ across that region. Since
$\partial_x\eta=-J_{\rm eff}/\mu(b)$, integration over~$\Omega$ gives~$
\Delta\eta = -J_{\rm eff}\int_\Omega dx/\mu(b(x))$. Therefore, we obtain
\begin{equation}
 J_{\rm eff}=-\mathcal{G}_\Omega\Delta\eta,
 \label{eq:drift_conductance}
\end{equation}
in which~$\mathcal{G}_\Omega = \left[
\int_\Omega dx/\mu(b(x))
\right]^{-1}$ is the effective conductance of the region.

For ripening, the perturbation variable is the mass imbalance~$\delta M$ between two neighboring aggregates with masses~$M+\delta M$ and~$M-\delta M$ (Fig.~\ref{fig:coarsening_patterns}), where the mass transfer between the two aggregates occurs through the low-density plateau separating them. In this region, the density is approximately constant at~$b_{-}$ and therefore the mobility is given by~$\mu(b) \simeq \mu(b_{-})$. For the considered stationary half-peak solution, the domain length is~$\Lambda/2$, so the effective conductance~$\mathcal{G}_\Omega$ for the low density plateau is given by 
\begin{align}
\mathcal{G}_{\rm plateau} &= \left[\int_{\rm plateau}\frac{dx}{\mu(b(x))} \right]^{-1}
\nonumber \\
&\simeq \left[\frac{\Lambda/2}{\mu(b_-)} \right]^{-1} = \frac{2\mu(b_-)}{\Lambda}. 
\label{eq:plateau_conductance}
\end{align}
The corresponding stationary value of the redistribution potential is denoted by~$\eta_{\rm stat}(M)$. Its mass derivative is estimated numerically~\cite{noteSI} from neighboring stationary profiles with masses~$M+\delta M$ and \(M-\delta M\): 
\begin{align}
\partial_M\eta_{\rm stat} \simeq
\frac{\eta_{\rm stat}(M+\delta M)
-\eta_{\rm stat}(M-\delta M)}{2\delta M} = \frac{\Delta \eta_{\rm rip}}{2 \delta M}. \label{eq:deriv_stat_eta}
\end{align}
Therefore, from Eqs.~(\ref{eq:drift_conductance}-\ref{eq:deriv_stat_eta}), we eventually find
\begin{align}
\frac{d}{dt}\delta M &=
J_{\rm rip} =-\mathcal{G}_{\rm plateau}\Delta\eta_{\rm rip} \nonumber \\ &= - \dfrac{4}{\Lambda} \mu(b_{-}) \partial_M \eta_{\text{stat}}\delta M = \sigma_{\rm rip} \delta M, \label{eq:growth_rate_ripening}
\end{align}
which leads to the ripening growth rate~\cite{weyer2025chemotaxis,weyer2025coarsening} in Eq.~(\ref{eq:sigma_rip}). 

For coalescence, the perturbation variable is the displacement of the peak, such that a small displacement~$\delta x$ changes the amount of mass on either side of the stationary peak. To leading order, this small mass difference~$\delta M$ is directly proportional to the difference in cell density between the peak and the surrounding low density plateau~\cite{weyer2025chemotaxis,weyer2025coarsening} and therefore given by~$\delta M_{\rm disp} \simeq (b_{+} - b_{-}) \delta x$, which induces a redistribution potential difference of 
\begin{align}
\Delta\eta_{\rm coal} \simeq 2\partial_M\eta_{\rm stat} (b_{+}-b_{-})\delta x. \label{eq:eta_coal}
\end{align}
Moreover, the effective conductance~$\mathcal{G}_{\Omega}$ in this case arises across the core of the dense peak itself unlike the low density plateau with constant density~$b_{-}$ in case of ripening. Therefore, it is given by the integral 
\begin{align}
	\mathcal{G}_{\rm peak} &= \left[\int_{\rm core}\frac{dx}{\mu(b(x))} \right]^{-1} \! .
\label{eq:peak_conductance}
\end{align}
The integral is evaluated numerically over the dense core of the mirrored stationary peak profile in the region satisfying~$b(x)>b_{-}+\frac{1}{2}(b_{+}-b_{-})$. Therefore, from Eqs.~(\ref{eq:eta_coal},\ref{eq:peak_conductance}), we can write 
\begin{align}
\frac{d}{dt}\delta x &=-\mathcal{G}_{\rm peak}\Delta\eta_{\rm coal} \nonumber \\ &= -2 \mathcal{G}_{\text{peak}}\left(b_{+} - b_{-}\right) \partial_M \eta_{\text{stat}} \delta x = \sigma_{\rm coal} \delta x, \label{eq:growth_rate_coalescence}
\end{align}
which leads to the coalescence growth rate in Eq.~(\ref{eq:sigma_coal}).

\clearpage
\detailtexcount{main}

\end{document}

%% file: main.bbl
%

%% file: main.bbl
\begin{thebibliography}{87}%
\makeatletter
\providecommand \@ifxundefined [1]{%
 \@ifx{#1\undefined}
}%
\providecommand \@ifnum [1]{%
 \ifnum #1\expandafter \@firstoftwo
 \else \expandafter \@secondoftwo
 \fi
}%
\providecommand \@ifx [1]{%
 \ifx #1\expandafter \@firstoftwo
 \else \expandafter \@secondoftwo
 \fi
}%
\providecommand \natexlab [1]{#1}%
\providecommand \enquote  [1]{``#1''}%
\providecommand \bibnamefont  [1]{#1}%
\providecommand \bibfnamefont [1]{#1}%
\providecommand \citenamefont [1]{#1}%
\providecommand \href@noop [0]{\@secondoftwo}%
\providecommand \href [0]{\begingroup \@sanitize@url \@href}%
\providecommand \@href[1]{\@@startlink{#1}\@@href}%
\providecommand \@@href[1]{\endgroup#1\@@endlink}%
\providecommand \@sanitize@url [0]{\catcode `\\12\catcode `\$12\catcode
  `\&12\catcode `\#12\catcode `\^12\catcode `\_12\catcode `\%12\relax}%
\providecommand \@@startlink[1]{}%
\providecommand \@@endlink[0]{}%
\providecommand \url  [0]{\begingroup\@sanitize@url \@url }%
\providecommand \@url [1]{\endgroup\@href {#1}{\urlprefix }}%
\providecommand \urlprefix  [0]{URL }%
\providecommand \Eprint [0]{\href }%
\providecommand \doibase [0]{https://doi.org/}%
\providecommand \selectlanguage [0]{\@gobble}%
\providecommand \bibinfo  [0]{\@secondoftwo}%
\providecommand \bibfield  [0]{\@secondoftwo}%
\providecommand \translation [1]{[#1]}%
\providecommand \BibitemOpen [0]{}%
\providecommand \bibitemStop [0]{}%
\providecommand \bibitemNoStop [0]{.\EOS\space}%
\providecommand \EOS [0]{\spacefactor3000\relax}%
\providecommand \BibitemShut  [1]{\csname bibitem#1\endcsname}%
\let\auto@bib@innerbib\@empty
\bibitem [{\citenamefont {Adler}(1975)}]{adler1975chemotaxis}%
  \BibitemOpen
  \bibfield  {author} {\bibinfo {author} {\bibfnamefont {J.}~\bibnamefont
  {Adler}},\ }\bibfield  {title} {\bibinfo {title} {Chemotaxis in bacteria},\
  }\href {https://doi.org/10.1146/annurev.bi.44.070175.002013} {\bibfield
  {journal} {\bibinfo  {journal} {Annu. Rev. Biochem.}\ }\textbf {\bibinfo
  {volume} {44}},\ \bibinfo {pages} {341} (\bibinfo {year} {1975})}\BibitemShut
  {NoStop}%
\bibitem [{\citenamefont {Berg}(1975)}]{berg1975chemotaxis}%
  \BibitemOpen
  \bibfield  {author} {\bibinfo {author} {\bibfnamefont {H.~C.}\ \bibnamefont
  {Berg}},\ }\bibfield  {title} {\bibinfo {title} {Chemotaxis in bacteria},\
  }\href {https://doi.org/10.1146/annurev.bb.04.060175.001003} {\bibfield
  {journal} {\bibinfo  {journal} {Annu. Rev. Biophys. Bioeng.}\ }\textbf
  {\bibinfo {volume} {4}},\ \bibinfo {pages} {119} (\bibinfo {year}
  {1975})}\BibitemShut {NoStop}%
\bibitem [{\citenamefont {Van~Haastert}\ and\ \citenamefont
  {Devreotes}(2004)}]{van2004chemotaxis}%
  \BibitemOpen
  \bibfield  {author} {\bibinfo {author} {\bibfnamefont {P.~J.~M.}\
  \bibnamefont {Van~Haastert}}\ and\ \bibinfo {author} {\bibfnamefont {P.~N.}\
  \bibnamefont {Devreotes}},\ }\bibfield  {title} {\bibinfo {title}
  {Chemotaxis: {{Signalling}} the way forward},\ }\href
  {https://doi.org/10.1038/nrm1435} {\bibfield  {journal} {\bibinfo  {journal}
  {Nat. Rev. Mol. Cell Biol.}\ }\textbf {\bibinfo {volume} {5}},\ \bibinfo
  {pages} {626} (\bibinfo {year} {2004})}\BibitemShut {NoStop}%
\bibitem [{\citenamefont {Swaney}\ \emph {et~al.}(2010)\citenamefont {Swaney},
  \citenamefont {Huang},\ and\ \citenamefont
  {Devreotes}}]{swaney2010eukaryotic}%
  \BibitemOpen
  \bibfield  {author} {\bibinfo {author} {\bibfnamefont {K.~F.}\ \bibnamefont
  {Swaney}}, \bibinfo {author} {\bibfnamefont {C.-H.}\ \bibnamefont {Huang}},\
  and\ \bibinfo {author} {\bibfnamefont {P.~N.}\ \bibnamefont {Devreotes}},\
  }\bibfield  {title} {\bibinfo {title} {Eukaryotic chemotaxis: {A} network of
  signaling pathways controls motility, directional sensing, and polarity},\
  }\href {https://doi.org/10.1146/annurev.biophys.093008.131228} {\bibfield
  {journal} {\bibinfo  {journal} {Annu. Rev. Biophys.}\ }\textbf {\bibinfo
  {volume} {39}},\ \bibinfo {pages} {265} (\bibinfo {year} {2010})}\BibitemShut
  {NoStop}%
\bibitem [{\citenamefont {Kolaczkowska}\ and\ \citenamefont
  {Kubes}(2013)}]{kolaczkowska2013neutrophil}%
  \BibitemOpen
  \bibfield  {author} {\bibinfo {author} {\bibfnamefont {E.}~\bibnamefont
  {Kolaczkowska}}\ and\ \bibinfo {author} {\bibfnamefont {P.}~\bibnamefont
  {Kubes}},\ }\bibfield  {title} {\bibinfo {title} {Neutrophil recruitment and
  function in health and inflammation},\ }\href
  {https://doi.org/10.1038/nri3399} {\bibfield  {journal} {\bibinfo  {journal}
  {Nat. Rev. Immunol.}\ }\textbf {\bibinfo {volume} {13}},\ \bibinfo {pages}
  {159} (\bibinfo {year} {2013})}\BibitemShut {NoStop}%
\bibitem [{\citenamefont {Berg}\ and\ \citenamefont
  {Brown}(1972)}]{berg1972chemotaxis}%
  \BibitemOpen
  \bibfield  {author} {\bibinfo {author} {\bibfnamefont {H.~C.}\ \bibnamefont
  {Berg}}\ and\ \bibinfo {author} {\bibfnamefont {D.~A.}\ \bibnamefont
  {Brown}},\ }\bibfield  {title} {\bibinfo {title} {Chemotaxis in
  \textit{{Escherichia} coli} analysed by three-dimensional tracking},\ }\href
  {https://doi.org/10.1038/239500a0} {\bibfield  {journal} {\bibinfo  {journal}
  {Nature}\ }\textbf {\bibinfo {volume} {239}},\ \bibinfo {pages} {500}
  (\bibinfo {year} {1972})}\BibitemShut {NoStop}%
\bibitem [{\citenamefont {Macnab}\ and\ \citenamefont
  {Koshland}(1972)}]{macnab1972}%
  \BibitemOpen
  \bibfield  {author} {\bibinfo {author} {\bibfnamefont {R.~M.}\ \bibnamefont
  {Macnab}}\ and\ \bibinfo {author} {\bibfnamefont {D.~E.}\ \bibnamefont
  {Koshland}},\ }\bibfield  {title} {\bibinfo {title} {The gradient-sensing
  mechanism in bacterial chemotaxis},\ }\href
  {https://doi.org/10.1073/pnas.69.9.2509} {\bibfield  {journal} {\bibinfo
  {journal} {Proc. Natl. Acad. Sci. USA}\ }\textbf {\bibinfo {volume} {69}},\
  \bibinfo {pages} {2509} (\bibinfo {year} {1972})}\BibitemShut {NoStop}%
\bibitem [{\citenamefont {Alirezaeizanjani}\ \emph {et~al.}(2020)\citenamefont
  {Alirezaeizanjani}, \citenamefont {Gro{\ss}mann}, \citenamefont {Pfeifer},
  \citenamefont {Hintsche},\ and\ \citenamefont {Beta}}]{Alirezaeizanjani2020}%
  \BibitemOpen
  \bibfield  {author} {\bibinfo {author} {\bibfnamefont {Z.}~\bibnamefont
  {Alirezaeizanjani}}, \bibinfo {author} {\bibfnamefont {R.}~\bibnamefont
  {Gro{\ss}mann}}, \bibinfo {author} {\bibfnamefont {V.}~\bibnamefont
  {Pfeifer}}, \bibinfo {author} {\bibfnamefont {M.}~\bibnamefont {Hintsche}},\
  and\ \bibinfo {author} {\bibfnamefont {C.}~\bibnamefont {Beta}},\ }\bibfield
  {title} {\bibinfo {title} {Chemotaxis strategies of bacteria with multiple
  run modes},\ }\href {https://doi.org/10.1126/sciadv.aaz6153} {\bibfield
  {journal} {\bibinfo  {journal} {Sci. Adv.}\ }\textbf {\bibinfo {volume}
  {6}},\ \bibinfo {pages} {eaaz6153} (\bibinfo {year} {2020})}\BibitemShut
  {NoStop}%
\bibitem [{\citenamefont {Pohl}\ \emph {et~al.}(2017)\citenamefont {Pohl},
  \citenamefont {Hintsche}, \citenamefont {Alirezaeizanjani}, \citenamefont
  {Seyrich}, \citenamefont {Beta},\ and\ \citenamefont {Stark}}]{Pohl2017}%
  \BibitemOpen
  \bibfield  {author} {\bibinfo {author} {\bibfnamefont {O.}~\bibnamefont
  {Pohl}}, \bibinfo {author} {\bibfnamefont {M.}~\bibnamefont {Hintsche}},
  \bibinfo {author} {\bibfnamefont {Z.}~\bibnamefont {Alirezaeizanjani}},
  \bibinfo {author} {\bibfnamefont {M.}~\bibnamefont {Seyrich}}, \bibinfo
  {author} {\bibfnamefont {C.}~\bibnamefont {Beta}},\ and\ \bibinfo {author}
  {\bibfnamefont {H.}~\bibnamefont {Stark}},\ }\bibfield  {title} {\bibinfo
  {title} {Inferring the chemotactic strategy of {{{\emph{P. putida}}}} and
  {{{\emph{E. coli}}}} using modified {{Kramers-Moyal}} coefficients},\ }\href
  {https://doi.org/10.1371/journal.pcbi.1005329} {\bibfield  {journal}
  {\bibinfo  {journal} {PLoS Comput. Biol.}\ }\textbf {\bibinfo {volume}
  {13}},\ \bibinfo {pages} {e1005329} (\bibinfo {year} {2017})}\BibitemShut
  {NoStop}%
\bibitem [{\citenamefont {Seyrich}\ \emph {et~al.}(2018)\citenamefont
  {Seyrich}, \citenamefont {Alirezaeizanjani}, \citenamefont {Beta},\ and\
  \citenamefont {Stark}}]{Seyrich2018}%
  \BibitemOpen
  \bibfield  {author} {\bibinfo {author} {\bibfnamefont {M.}~\bibnamefont
  {Seyrich}}, \bibinfo {author} {\bibfnamefont {Z.}~\bibnamefont
  {Alirezaeizanjani}}, \bibinfo {author} {\bibfnamefont {C.}~\bibnamefont
  {Beta}},\ and\ \bibinfo {author} {\bibfnamefont {H.}~\bibnamefont {Stark}},\
  }\bibfield  {title} {\bibinfo {title} {Statistical parameter inference of
  bacterial swimming strategies},\ }\href
  {https://doi.org/10.1088/1367-2630/aae72c} {\bibfield  {journal} {\bibinfo
  {journal} {New J. Phys.}\ }\textbf {\bibinfo {volume} {20}},\ \bibinfo
  {pages} {103033} (\bibinfo {year} {2018})}\BibitemShut {NoStop}%
\bibitem [{\citenamefont {G\'{o}mez~Nava}\ \emph {et~al.}(2020)\citenamefont
  {G\'{o}mez~Nava}, \citenamefont {Gro{\ss{}}mann}, \citenamefont {Hintsche},
  \citenamefont {Beta},\ and\ \citenamefont {Peruani}}]{gomeznava2020novel}%
  \BibitemOpen
  \bibfield  {author} {\bibinfo {author} {\bibfnamefont {L.}~\bibnamefont
  {G\'{o}mez~Nava}}, \bibinfo {author} {\bibfnamefont {R.}~\bibnamefont
  {Gro{\ss{}}mann}}, \bibinfo {author} {\bibfnamefont {M.}~\bibnamefont
  {Hintsche}}, \bibinfo {author} {\bibfnamefont {C.}~\bibnamefont {Beta}},\
  and\ \bibinfo {author} {\bibfnamefont {F.}~\bibnamefont {Peruani}},\
  }\bibfield  {title} {\bibinfo {title} {A novel approach to chemotaxis:
  {A}ctive particles guided by internal clocks},\ }\href
  {https://doi.org/10.1209/0295-5075/130/68002} {\bibfield  {journal} {\bibinfo
   {journal} {Europhys. Lett.}\ }\textbf {\bibinfo {volume} {130}},\ \bibinfo
  {pages} {68002} (\bibinfo {year} {2020})}\BibitemShut {NoStop}%
\bibitem [{\citenamefont {Seyrich}\ \emph {et~al.}(2019)\citenamefont
  {Seyrich}, \citenamefont {Palugniok},\ and\ \citenamefont
  {Stark}}]{seyrich2019traveling}%
  \BibitemOpen
  \bibfield  {author} {\bibinfo {author} {\bibfnamefont {M.}~\bibnamefont
  {Seyrich}}, \bibinfo {author} {\bibfnamefont {A.}~\bibnamefont {Palugniok}},\
  and\ \bibinfo {author} {\bibfnamefont {H.}~\bibnamefont {Stark}},\ }\bibfield
   {title} {\bibinfo {title} {Traveling concentration pulses of bacteria in a
  generalized {{Keller}}-{{Segel}} model},\ }\href
  {https://doi.org/10.1088/1367-2630/ab4522} {\bibfield  {journal} {\bibinfo
  {journal} {New J. Phys.}\ }\textbf {\bibinfo {volume} {21}},\ \bibinfo
  {pages} {103001} (\bibinfo {year} {2019})}\BibitemShut {NoStop}%
\bibitem [{\citenamefont {Nakai}\ \emph {et~al.}(2021)\citenamefont {Nakai},
  \citenamefont {Ando},\ and\ \citenamefont {Goto}}]{Nakai2021}%
  \BibitemOpen
  \bibfield  {author} {\bibinfo {author} {\bibfnamefont {T.}~\bibnamefont
  {Nakai}}, \bibinfo {author} {\bibfnamefont {T.}~\bibnamefont {Ando}},\ and\
  \bibinfo {author} {\bibfnamefont {T.}~\bibnamefont {Goto}},\ }\bibfield
  {title} {\bibinfo {title} {Biased reorientation in the chemotaxis of
  peritrichous bacteria {{{\emph{Salmonella enterica}}}} serovar
  {{{Typhimurium}}}},\ }\href {https://doi.org/10.1016/j.bpj.2021.04.033}
  {\bibfield  {journal} {\bibinfo  {journal} {Biophys. J.}\ }\textbf {\bibinfo
  {volume} {120}},\ \bibinfo {pages} {2623} (\bibinfo {year}
  {2021})}\BibitemShut {NoStop}%
\bibitem [{\citenamefont {Bhattacharjee}\ \emph {et~al.}(2021)\citenamefont
  {Bhattacharjee}, \citenamefont {Amchin}, \citenamefont {Ott}, \citenamefont
  {Kratz},\ and\ \citenamefont {Datta}}]{Bhattacharjee2021}%
  \BibitemOpen
  \bibfield  {author} {\bibinfo {author} {\bibfnamefont {T.}~\bibnamefont
  {Bhattacharjee}}, \bibinfo {author} {\bibfnamefont {D.~B.}\ \bibnamefont
  {Amchin}}, \bibinfo {author} {\bibfnamefont {J.~A.}\ \bibnamefont {Ott}},
  \bibinfo {author} {\bibfnamefont {F.}~\bibnamefont {Kratz}},\ and\ \bibinfo
  {author} {\bibfnamefont {S.~S.}\ \bibnamefont {Datta}},\ }\bibfield  {title}
  {\bibinfo {title} {Chemotactic migration of bacteria in porous media},\
  }\href {https://doi.org/10.1016/j.bpj.2021.05.012} {\bibfield  {journal}
  {\bibinfo  {journal} {Biophys. J.}\ }\textbf {\bibinfo {volume} {120}},\
  \bibinfo {pages} {3483} (\bibinfo {year} {2021})}\BibitemShut {NoStop}%
\bibitem [{\citenamefont {Beier}\ \emph {et~al.}(2026)\citenamefont {Beier},
  \citenamefont {Pfeifer}, \citenamefont {Datta}, \citenamefont
  {Gro{\ss}mann},\ and\ \citenamefont {Beta}}]{beier2026dual}%
  \BibitemOpen
  \bibfield  {author} {\bibinfo {author} {\bibfnamefont {S.}~\bibnamefont
  {Beier}}, \bibinfo {author} {\bibfnamefont {V.}~\bibnamefont {Pfeifer}},
  \bibinfo {author} {\bibfnamefont {A.}~\bibnamefont {Datta}}, \bibinfo
  {author} {\bibfnamefont {R.}~\bibnamefont {Gro{\ss}mann}},\ and\ \bibinfo
  {author} {\bibfnamefont {C.}~\bibnamefont {Beta}},\ }\bibfield  {title}
  {\bibinfo {title} {Dual chemotaxis strategy of bacteria in porous media:
  {{Run}}-time and turn-angle bias},\ }\href
  {https://doi.org/10.1016/j.bpj.2026.06.002} {\bibfield  {journal} {\bibinfo
  {journal} {Biophys. J.}\ }\textbf {\bibinfo {volume} {125}},\ \bibinfo
  {pages} {3570} (\bibinfo {year} {2026})}\BibitemShut {NoStop}%
\bibitem [{\citenamefont {Adler}(1966)}]{adler1966chemotaxis}%
  \BibitemOpen
  \bibfield  {author} {\bibinfo {author} {\bibfnamefont {J.}~\bibnamefont
  {Adler}},\ }\bibfield  {title} {\bibinfo {title} {Chemotaxis in bacteria:
  {{Motile}} {{{\emph{Escherichia coli}}}} migrate in bands that are influenced
  by oxygen and organic nutrients},\ }\href
  {https://doi.org/10.1126/science.153.3737.708} {\bibfield  {journal}
  {\bibinfo  {journal} {Science}\ }\textbf {\bibinfo {volume} {153}},\ \bibinfo
  {pages} {708} (\bibinfo {year} {1966})}\BibitemShut {NoStop}%
\bibitem [{\citenamefont {Grigson}\ \emph {et~al.}(2026)\citenamefont
  {Grigson}, \citenamefont {Hutton}, \citenamefont {{Carlson-Jones}},
  \citenamefont {Paterson}, \citenamefont {Giles}, \citenamefont {Harker},
  \citenamefont {Grigson}, \citenamefont {Annells}, \citenamefont {Roach},
  \citenamefont {Speck} \emph {et~al.}}]{grigson2026co}%
  \BibitemOpen
  \bibfield  {author} {\bibinfo {author} {\bibfnamefont {S.~R.}\ \bibnamefont
  {Grigson}}, \bibinfo {author} {\bibfnamefont {A.~L.}\ \bibnamefont {Hutton}},
  \bibinfo {author} {\bibfnamefont {J.~A.}\ \bibnamefont {{Carlson-Jones}}},
  \bibinfo {author} {\bibfnamefont {J.~S.}\ \bibnamefont {Paterson}}, \bibinfo
  {author} {\bibfnamefont {S.~K.}\ \bibnamefont {Giles}}, \bibinfo {author}
  {\bibfnamefont {C.~M.}\ \bibnamefont {Harker}}, \bibinfo {author}
  {\bibfnamefont {A.~K.}\ \bibnamefont {Grigson}}, \bibinfo {author}
  {\bibfnamefont {A.~N.}\ \bibnamefont {Annells}}, \bibinfo {author}
  {\bibfnamefont {M.~J.}\ \bibnamefont {Roach}}, \bibinfo {author}
  {\bibfnamefont {P.~G.}\ \bibnamefont {Speck}}, \emph {et~al.},\ }\bibfield
  {title} {\bibinfo {title} {Co-migration of hundreds of microbial species over
  metres drives selection and promotes non-motile hitchhikers},\ }\href
  {https://doi.org/10.1038/s41467-026-72998-1} {\bibfield  {journal} {\bibinfo
  {journal} {Nat. Commun.}\ }\textbf {\bibinfo {volume} {17}},\ \bibinfo
  {pages} {6640} (\bibinfo {year} {2026})}\BibitemShut {NoStop}%
\bibitem [{\citenamefont {Budrene}\ and\ \citenamefont
  {Berg}(1991)}]{budrene1991complex}%
  \BibitemOpen
  \bibfield  {author} {\bibinfo {author} {\bibfnamefont {E.~O.}\ \bibnamefont
  {Budrene}}\ and\ \bibinfo {author} {\bibfnamefont {H.~C.}\ \bibnamefont
  {Berg}},\ }\bibfield  {title} {\bibinfo {title} {Complex patterns formed by
  motile cells of \textit{{Escherichia} Coli}},\ }\href
  {https://doi.org/10.1038/349630a0} {\bibfield  {journal} {\bibinfo  {journal}
  {Nature}\ }\textbf {\bibinfo {volume} {349}},\ \bibinfo {pages} {630}
  (\bibinfo {year} {1991})}\BibitemShut {NoStop}%
\bibitem [{\citenamefont {Budrene}\ and\ \citenamefont
  {Berg}(1995)}]{budrene1995dynamics}%
  \BibitemOpen
  \bibfield  {author} {\bibinfo {author} {\bibfnamefont {E.~O.}\ \bibnamefont
  {Budrene}}\ and\ \bibinfo {author} {\bibfnamefont {H.~C.}\ \bibnamefont
  {Berg}},\ }\bibfield  {title} {\bibinfo {title} {Dynamics of formation of
  symmetrical patterns by chemotactic bacteria},\ }\href
  {https://doi.org/10.1038/376049a0} {\bibfield  {journal} {\bibinfo  {journal}
  {Nature}\ }\textbf {\bibinfo {volume} {376}},\ \bibinfo {pages} {49}
  (\bibinfo {year} {1995})}\BibitemShut {NoStop}%
\bibitem [{\citenamefont {Woodward}\ \emph {et~al.}(1995)\citenamefont
  {Woodward}, \citenamefont {Tyson}, \citenamefont {Myerscough}, \citenamefont
  {Murray}, \citenamefont {Budrene},\ and\ \citenamefont
  {Berg}}]{woodward1995spatio}%
  \BibitemOpen
  \bibfield  {author} {\bibinfo {author} {\bibfnamefont {D.~E.}\ \bibnamefont
  {Woodward}}, \bibinfo {author} {\bibfnamefont {R.}~\bibnamefont {Tyson}},
  \bibinfo {author} {\bibfnamefont {M.~R.}\ \bibnamefont {Myerscough}},
  \bibinfo {author} {\bibfnamefont {J.~D.}\ \bibnamefont {Murray}}, \bibinfo
  {author} {\bibfnamefont {E.~O.}\ \bibnamefont {Budrene}},\ and\ \bibinfo
  {author} {\bibfnamefont {H.~C.}\ \bibnamefont {Berg}},\ }\bibfield  {title}
  {\bibinfo {title} {Spatio-temporal patterns generated by
  \textit{{Salmonella}} {T}yphimurium},\ }\href
  {https://doi.org/10.1016/s0006-3495(95)80400-5} {\bibfield  {journal}
  {\bibinfo  {journal} {Biophys. J.}\ }\textbf {\bibinfo {volume} {68}},\
  \bibinfo {pages} {2181} (\bibinfo {year} {1995})}\BibitemShut {NoStop}%
\bibitem [{\citenamefont {Blat}\ and\ \citenamefont
  {Eisenbach}(1995)}]{blat1995tar}%
  \BibitemOpen
  \bibfield  {author} {\bibinfo {author} {\bibfnamefont {Y.}~\bibnamefont
  {Blat}}\ and\ \bibinfo {author} {\bibfnamefont {M.}~\bibnamefont
  {Eisenbach}},\ }\bibfield  {title} {\bibinfo {title} {Tar-dependent
  and-independent pattern formation by \textit{{Salmonella}} {T}yphimurium},\
  }\href {https://doi.org/10.1128/jb.177.7.1683-1691.1995} {\bibfield
  {journal} {\bibinfo  {journal} {J. Bacteriol.}\ }\textbf {\bibinfo {volume}
  {177}},\ \bibinfo {pages} {1683} (\bibinfo {year} {1995})}\BibitemShut
  {NoStop}%
\bibitem [{\citenamefont {Fisher}\ \emph {et~al.}(1989)\citenamefont {Fisher},
  \citenamefont {Merkl},\ and\ \citenamefont
  {Gerisch}}]{fisher1989quantitative}%
  \BibitemOpen
  \bibfield  {author} {\bibinfo {author} {\bibfnamefont {P.~R.}\ \bibnamefont
  {Fisher}}, \bibinfo {author} {\bibfnamefont {R.}~\bibnamefont {Merkl}},\ and\
  \bibinfo {author} {\bibfnamefont {G.}~\bibnamefont {Gerisch}},\ }\bibfield
  {title} {\bibinfo {title} {Quantitative analysis of cell motility and
  chemotaxis in \textit{{Dictyostelium} discoideum} by using an image
  processing system and a novel chemotaxis chamber providing stationary
  chemical gradients.},\ }\href {https://doi.org/10.1083/jcb.108.3.973}
  {\bibfield  {journal} {\bibinfo  {journal} {J. Cell Biol.}\ }\textbf
  {\bibinfo {volume} {108}},\ \bibinfo {pages} {973} (\bibinfo {year}
  {1989})}\BibitemShut {NoStop}%
\bibitem [{\citenamefont {Song}\ \emph {et~al.}(2006)\citenamefont {Song},
  \citenamefont {Nadkarni}, \citenamefont {B{\"o}deker}, \citenamefont {Beta},
  \citenamefont {Bae}, \citenamefont {Franck}, \citenamefont {Rappel},
  \citenamefont {Loomis},\ and\ \citenamefont
  {Bodenschatz}}]{song2006dictyostelium}%
  \BibitemOpen
  \bibfield  {author} {\bibinfo {author} {\bibfnamefont {L.}~\bibnamefont
  {Song}}, \bibinfo {author} {\bibfnamefont {S.~M.}\ \bibnamefont {Nadkarni}},
  \bibinfo {author} {\bibfnamefont {H.~U.}\ \bibnamefont {B{\"o}deker}},
  \bibinfo {author} {\bibfnamefont {C.}~\bibnamefont {Beta}}, \bibinfo {author}
  {\bibfnamefont {A.}~\bibnamefont {Bae}}, \bibinfo {author} {\bibfnamefont
  {C.}~\bibnamefont {Franck}}, \bibinfo {author} {\bibfnamefont {W.-J.}\
  \bibnamefont {Rappel}}, \bibinfo {author} {\bibfnamefont {W.~F.}\
  \bibnamefont {Loomis}},\ and\ \bibinfo {author} {\bibfnamefont
  {E.}~\bibnamefont {Bodenschatz}},\ }\bibfield  {title} {\bibinfo {title}
  {Dictyostelium discoideum chemotaxis: {T}hreshold for directed motion},\
  }\href {https://doi.org/10.1016/j.ejcb.2006.01.012} {\bibfield  {journal}
  {\bibinfo  {journal} {Eur. J. Cell Biol.}\ }\textbf {\bibinfo {volume}
  {85}},\ \bibinfo {pages} {981} (\bibinfo {year} {2006})}\BibitemShut
  {NoStop}%
\bibitem [{\citenamefont {Nichols}\ \emph {et~al.}(2015)\citenamefont
  {Nichols}, \citenamefont {Veltman},\ and\ \citenamefont
  {Kay}}]{nichols2015chemotaxis}%
  \BibitemOpen
  \bibfield  {author} {\bibinfo {author} {\bibfnamefont {J.~M.}\ \bibnamefont
  {Nichols}}, \bibinfo {author} {\bibfnamefont {D.}~\bibnamefont {Veltman}},\
  and\ \bibinfo {author} {\bibfnamefont {R.~R.}\ \bibnamefont {Kay}},\
  }\bibfield  {title} {\bibinfo {title} {Chemotaxis of a model organism:
  {P}rogress with \textit{{Dictyostelium}}},\ }\href
  {https://doi.org/10.1016/j.ceb.2015.06.005} {\bibfield  {journal} {\bibinfo
  {journal} {Curr. Opin. Cell Biol.}\ }\textbf {\bibinfo {volume} {36}},\
  \bibinfo {pages} {7} (\bibinfo {year} {2015})}\BibitemShut {NoStop}%
\bibitem [{\citenamefont {Park}\ \emph {et~al.}(2003)\citenamefont {Park},
  \citenamefont {Wolanin}, \citenamefont {Yuzbashyan}, \citenamefont {Lin},
  \citenamefont {Darnton}, \citenamefont {Stock}, \citenamefont {Silberzan},\
  and\ \citenamefont {Austin}}]{park2003influence}%
  \BibitemOpen
  \bibfield  {author} {\bibinfo {author} {\bibfnamefont {S.}~\bibnamefont
  {Park}}, \bibinfo {author} {\bibfnamefont {P.~M.}\ \bibnamefont {Wolanin}},
  \bibinfo {author} {\bibfnamefont {E.~A.}\ \bibnamefont {Yuzbashyan}},
  \bibinfo {author} {\bibfnamefont {H.}~\bibnamefont {Lin}}, \bibinfo {author}
  {\bibfnamefont {N.~C.}\ \bibnamefont {Darnton}}, \bibinfo {author}
  {\bibfnamefont {J.~B.}\ \bibnamefont {Stock}}, \bibinfo {author}
  {\bibfnamefont {P.}~\bibnamefont {Silberzan}},\ and\ \bibinfo {author}
  {\bibfnamefont {R.}~\bibnamefont {Austin}},\ }\bibfield  {title} {\bibinfo
  {title} {Influence of topology on bacterial social interaction},\ }\href
  {https://doi.org/10.1073/pnas.1935975100} {\bibfield  {journal} {\bibinfo
  {journal} {Proc. Natl. Acad. Sci. USA}\ }\textbf {\bibinfo {volume} {100}},\
  \bibinfo {pages} {13910} (\bibinfo {year} {2003})}\BibitemShut {NoStop}%
\bibitem [{\citenamefont {Scheidweiler}\ \emph {et~al.}(2024)\citenamefont
  {Scheidweiler}, \citenamefont {Bordoloi}, \citenamefont {Jiao}, \citenamefont
  {Sentchilo}, \citenamefont {Bollani}, \citenamefont {Chhun}, \citenamefont
  {Engel},\ and\ \citenamefont {De~Anna}}]{scheidweiler2024spatial}%
  \BibitemOpen
  \bibfield  {author} {\bibinfo {author} {\bibfnamefont {D.}~\bibnamefont
  {Scheidweiler}}, \bibinfo {author} {\bibfnamefont {A.~D.}\ \bibnamefont
  {Bordoloi}}, \bibinfo {author} {\bibfnamefont {W.}~\bibnamefont {Jiao}},
  \bibinfo {author} {\bibfnamefont {V.}~\bibnamefont {Sentchilo}}, \bibinfo
  {author} {\bibfnamefont {M.}~\bibnamefont {Bollani}}, \bibinfo {author}
  {\bibfnamefont {A.}~\bibnamefont {Chhun}}, \bibinfo {author} {\bibfnamefont
  {P.}~\bibnamefont {Engel}},\ and\ \bibinfo {author} {\bibfnamefont
  {P.}~\bibnamefont {De~Anna}},\ }\bibfield  {title} {\bibinfo {title} {Spatial
  structure, chemotaxis and quorum sensing shape bacterial biomass accumulation
  in complex porous media},\ }\href
  {https://doi.org/10.1038/s41467-023-44267-y} {\bibfield  {journal} {\bibinfo
  {journal} {Nat. Commun.}\ }\textbf {\bibinfo {volume} {15}},\ \bibinfo
  {pages} {191} (\bibinfo {year} {2024})}\BibitemShut {NoStop}%
\bibitem [{\citenamefont {Phan}\ \emph {et~al.}(2024)\citenamefont {Phan},
  \citenamefont {Mattingly}, \citenamefont {Vo}, \citenamefont {Marvin},
  \citenamefont {Looger},\ and\ \citenamefont {Emonet}}]{phan2024direct}%
  \BibitemOpen
  \bibfield  {author} {\bibinfo {author} {\bibfnamefont {T.~V.}\ \bibnamefont
  {Phan}}, \bibinfo {author} {\bibfnamefont {H.~H.}\ \bibnamefont {Mattingly}},
  \bibinfo {author} {\bibfnamefont {L.}~\bibnamefont {Vo}}, \bibinfo {author}
  {\bibfnamefont {J.~S.}\ \bibnamefont {Marvin}}, \bibinfo {author}
  {\bibfnamefont {L.~L.}\ \bibnamefont {Looger}},\ and\ \bibinfo {author}
  {\bibfnamefont {T.}~\bibnamefont {Emonet}},\ }\bibfield  {title} {\bibinfo
  {title} {Direct measurement of dynamic attractant gradients reveals breakdown
  of the {{Patlak}}-{{Keller}}-{{Segel}} chemotaxis model},\ }\href
  {https://doi.org/10.1073/pnas.2309251121} {\bibfield  {journal} {\bibinfo
  {journal} {Proc. Natl. Acad. Sci. USA}\ }\textbf {\bibinfo {volume} {121}},\
  \bibinfo {pages} {e2309251121} (\bibinfo {year} {2024})}\BibitemShut
  {NoStop}%
\bibitem [{\citenamefont {Keller}\ and\ \citenamefont
  {Segel}(1970)}]{keller1970initiation}%
  \BibitemOpen
  \bibfield  {author} {\bibinfo {author} {\bibfnamefont {E.~F.}\ \bibnamefont
  {Keller}}\ and\ \bibinfo {author} {\bibfnamefont {L.~A.}\ \bibnamefont
  {Segel}},\ }\bibfield  {title} {\bibinfo {title} {Initiation of slime mold
  aggregation viewed as an instability},\ }\href
  {https://doi.org/10.1016/0022-5193(70)90092-5} {\bibfield  {journal}
  {\bibinfo  {journal} {J. Theor. Biol.}\ }\textbf {\bibinfo {volume} {26}},\
  \bibinfo {pages} {399} (\bibinfo {year} {1970})}\BibitemShut {NoStop}%
\bibitem [{\citenamefont {Keller}\ and\ \citenamefont
  {Segel}(1971{\natexlab{a}})}]{keller1971model}%
  \BibitemOpen
  \bibfield  {author} {\bibinfo {author} {\bibfnamefont {E.~F.}\ \bibnamefont
  {Keller}}\ and\ \bibinfo {author} {\bibfnamefont {L.~A.}\ \bibnamefont
  {Segel}},\ }\bibfield  {title} {\bibinfo {title} {Model for chemotaxis},\
  }\href {https://doi.org/10.1016/0022-5193(71)90050-6} {\bibfield  {journal}
  {\bibinfo  {journal} {J. Theor. Biol.}\ }\textbf {\bibinfo {volume} {30}},\
  \bibinfo {pages} {225} (\bibinfo {year} {1971}{\natexlab{a}})}\BibitemShut
  {NoStop}%
\bibitem [{\citenamefont {Keller}\ and\ \citenamefont
  {Segel}(1971{\natexlab{b}})}]{keller1971traveling}%
  \BibitemOpen
  \bibfield  {author} {\bibinfo {author} {\bibfnamefont {E.~F.}\ \bibnamefont
  {Keller}}\ and\ \bibinfo {author} {\bibfnamefont {L.~A.}\ \bibnamefont
  {Segel}},\ }\bibfield  {title} {\bibinfo {title} {Traveling bands of
  chemotactic bacteria: {A} theoretical analysis},\ }\href
  {https://doi.org/10.1016/0022-5193(71)90051-8} {\bibfield  {journal}
  {\bibinfo  {journal} {J. Theor. Biol.}\ }\textbf {\bibinfo {volume} {30}},\
  \bibinfo {pages} {235} (\bibinfo {year} {1971}{\natexlab{b}})}\BibitemShut
  {NoStop}%
\bibitem [{\citenamefont {Brenner}\ \emph {et~al.}(1998)\citenamefont
  {Brenner}, \citenamefont {Levitov},\ and\ \citenamefont
  {Budrene}}]{brenner1998physical}%
  \BibitemOpen
  \bibfield  {author} {\bibinfo {author} {\bibfnamefont {M.~P.}\ \bibnamefont
  {Brenner}}, \bibinfo {author} {\bibfnamefont {L.~S.}\ \bibnamefont
  {Levitov}},\ and\ \bibinfo {author} {\bibfnamefont {E.~O.}\ \bibnamefont
  {Budrene}},\ }\bibfield  {title} {\bibinfo {title} {Physical mechanisms for
  chemotactic pattern formation by bacteria},\ }\href
  {https://doi.org/10.1016/s0006-3495(98)77880-4} {\bibfield  {journal}
  {\bibinfo  {journal} {Biophys. J.}\ }\textbf {\bibinfo {volume} {74}},\
  \bibinfo {pages} {1677} (\bibinfo {year} {1998})}\BibitemShut {NoStop}%
\bibitem [{\citenamefont {Tyson}\ \emph {et~al.}(1999)\citenamefont {Tyson},
  \citenamefont {Lubkin},\ and\ \citenamefont {Murray}}]{tyson1999minimal}%
  \BibitemOpen
  \bibfield  {author} {\bibinfo {author} {\bibfnamefont {R.}~\bibnamefont
  {Tyson}}, \bibinfo {author} {\bibfnamefont {S.~R.}\ \bibnamefont {Lubkin}},\
  and\ \bibinfo {author} {\bibfnamefont {J.~D.}\ \bibnamefont {Murray}},\
  }\bibfield  {title} {\bibinfo {title} {A minimal mechanism for bacterial
  pattern formation},\ }\href {https://doi.org/10.1098/rspb.1999.0637}
  {\bibfield  {journal} {\bibinfo  {journal} {Proc. R. Soc. B: Biol. Sci.}\
  }\textbf {\bibinfo {volume} {266}},\ \bibinfo {pages} {299} (\bibinfo {year}
  {1999})}\BibitemShut {NoStop}%
\bibitem [{\citenamefont {Ma}\ and\ \citenamefont
  {Wang}(2014)}]{ma2014dynamic}%
  \BibitemOpen
  \bibfield  {author} {\bibinfo {author} {\bibfnamefont {T.}~\bibnamefont
  {Ma}}\ and\ \bibinfo {author} {\bibfnamefont {S.}~\bibnamefont {Wang}},\
  }\bibfield  {title} {\bibinfo {title} {Dynamic transition and pattern
  formation for chemotactic systems},\ }\href
  {https://doi.org/10.3934/dcdsb.2014.19.2809} {\bibfield  {journal} {\bibinfo
  {journal} {Discrete Contin. Dyn. Syst.--B}\ }\textbf {\bibinfo {volume}
  {19}},\ \bibinfo {pages} {2809} (\bibinfo {year} {2014})}\BibitemShut
  {NoStop}%
\bibitem [{\citenamefont {Cremer}\ \emph {et~al.}(2019)\citenamefont {Cremer},
  \citenamefont {Honda}, \citenamefont {Tang}, \citenamefont {{Wong-Ng}},
  \citenamefont {Vergassola},\ and\ \citenamefont
  {Hwa}}]{cremer2019chemotaxis}%
  \BibitemOpen
  \bibfield  {author} {\bibinfo {author} {\bibfnamefont {J.}~\bibnamefont
  {Cremer}}, \bibinfo {author} {\bibfnamefont {T.}~\bibnamefont {Honda}},
  \bibinfo {author} {\bibfnamefont {Y.}~\bibnamefont {Tang}}, \bibinfo {author}
  {\bibfnamefont {J.}~\bibnamefont {{Wong-Ng}}}, \bibinfo {author}
  {\bibfnamefont {M.}~\bibnamefont {Vergassola}},\ and\ \bibinfo {author}
  {\bibfnamefont {T.}~\bibnamefont {Hwa}},\ }\bibfield  {title} {\bibinfo
  {title} {Chemotaxis as a navigation strategy to boost range expansion},\
  }\href {https://doi.org/10.1038/s41586-019-1733-y} {\bibfield  {journal}
  {\bibinfo  {journal} {Nature}\ }\textbf {\bibinfo {volume} {575}},\ \bibinfo
  {pages} {658} (\bibinfo {year} {2019})}\BibitemShut {NoStop}%
\bibitem [{\citenamefont {Zhao}\ \emph {et~al.}(2023)\citenamefont {Zhao},
  \citenamefont {Ko{\v s}mrlj},\ and\ \citenamefont
  {Datta}}]{zhao2023chemotactic}%
  \BibitemOpen
  \bibfield  {author} {\bibinfo {author} {\bibfnamefont {H.}~\bibnamefont
  {Zhao}}, \bibinfo {author} {\bibfnamefont {A.}~\bibnamefont {Ko{\v s}mrlj}},\
  and\ \bibinfo {author} {\bibfnamefont {S.~S.}\ \bibnamefont {Datta}},\
  }\bibfield  {title} {\bibinfo {title} {Chemotactic motility-induced phase
  separation},\ }\href {https://doi.org/10.1103/physrevlett.131.118301}
  {\bibfield  {journal} {\bibinfo  {journal} {Phys. Rev. Lett.}\ }\textbf
  {\bibinfo {volume} {131}},\ \bibinfo {pages} {118301} (\bibinfo {year}
  {2023})}\BibitemShut {NoStop}%
\bibitem [{\citenamefont {Kwon}\ \emph {et~al.}(2025)\citenamefont {Kwon},
  \citenamefont {Oh},\ and\ \citenamefont {Baek}}]{kwon2025phase}%
  \BibitemOpen
  \bibfield  {author} {\bibinfo {author} {\bibfnamefont {E.}~\bibnamefont
  {Kwon}}, \bibinfo {author} {\bibfnamefont {Y.}~\bibnamefont {Oh}},\ and\
  \bibinfo {author} {\bibfnamefont {Y.}~\bibnamefont {Baek}},\ }\bibfield
  {title} {\bibinfo {title} {Phase separation of chemokinetic active
  particles},\ }\href {https://doi.org/10.1038/s42005-025-02211-0} {\bibfield
  {journal} {\bibinfo  {journal} {Commun. Phys.}\ }\textbf {\bibinfo {volume}
  {8}},\ \bibinfo {pages} {288} (\bibinfo {year} {2025})}\BibitemShut {NoStop}%
\bibitem [{\citenamefont {Ra{\ss}hofer}\ \emph {et~al.}(2025)\citenamefont
  {Ra{\ss}hofer}, \citenamefont {Bauer}, \citenamefont {Ziepke}, \citenamefont
  {Maryshev},\ and\ \citenamefont {Frey}}]{rasshofer2025capillary}%
  \BibitemOpen
  \bibfield  {author} {\bibinfo {author} {\bibfnamefont {F.}~\bibnamefont
  {Ra{\ss}hofer}}, \bibinfo {author} {\bibfnamefont {S.}~\bibnamefont {Bauer}},
  \bibinfo {author} {\bibfnamefont {A.}~\bibnamefont {Ziepke}}, \bibinfo
  {author} {\bibfnamefont {I.}~\bibnamefont {Maryshev}},\ and\ \bibinfo
  {author} {\bibfnamefont {E.}~\bibnamefont {Frey}},\ }\bibfield  {title}
  {\bibinfo {title} {Capillary wave formation in conserved active emulsions},\
  }\href {https://doi.org/10.1103/lqws-dkmy} {\bibfield  {journal} {\bibinfo
  {journal} {Phys. Rev. Research}\ }\textbf {\bibinfo {volume} {7}},\ \bibinfo
  {pages} {043240} (\bibinfo {year} {2025})}\BibitemShut {NoStop}%
\bibitem [{\citenamefont {Grauer}\ \emph {et~al.}(2020)\citenamefont {Grauer},
  \citenamefont {L{\"o}wen}, \citenamefont {Be'er},\ and\ \citenamefont
  {Liebchen}}]{grauer2020swarm}%
  \BibitemOpen
  \bibfield  {author} {\bibinfo {author} {\bibfnamefont {J.}~\bibnamefont
  {Grauer}}, \bibinfo {author} {\bibfnamefont {H.}~\bibnamefont {L{\"o}wen}},
  \bibinfo {author} {\bibfnamefont {A.}~\bibnamefont {Be'er}},\ and\ \bibinfo
  {author} {\bibfnamefont {B.}~\bibnamefont {Liebchen}},\ }\bibfield  {title}
  {\bibinfo {title} {Swarm hunting and cluster ejections in chemically
  communicating active mixtures},\ }\href
  {https://doi.org/10.1038/s41598-020-62324-0} {\bibfield  {journal} {\bibinfo
  {journal} {Sci. Rep.}\ }\textbf {\bibinfo {volume} {10}},\ \bibinfo {pages}
  {5594} (\bibinfo {year} {2020})}\BibitemShut {NoStop}%
\bibitem [{\citenamefont {Dinelli}\ \emph {et~al.}(2023)\citenamefont
  {Dinelli}, \citenamefont {O'Byrne}, \citenamefont {Curatolo}, \citenamefont
  {Zhao}, \citenamefont {Sollich},\ and\ \citenamefont
  {Tailleur}}]{dinelli2023non}%
  \BibitemOpen
  \bibfield  {author} {\bibinfo {author} {\bibfnamefont {A.}~\bibnamefont
  {Dinelli}}, \bibinfo {author} {\bibfnamefont {J.}~\bibnamefont {O'Byrne}},
  \bibinfo {author} {\bibfnamefont {A.}~\bibnamefont {Curatolo}}, \bibinfo
  {author} {\bibfnamefont {Y.}~\bibnamefont {Zhao}}, \bibinfo {author}
  {\bibfnamefont {P.}~\bibnamefont {Sollich}},\ and\ \bibinfo {author}
  {\bibfnamefont {J.}~\bibnamefont {Tailleur}},\ }\bibfield  {title} {\bibinfo
  {title} {Non-reciprocity across scales in active mixtures},\ }\href
  {https://doi.org/10.1038/s41467-023-42713-5} {\bibfield  {journal} {\bibinfo
  {journal} {Nat. Commun.}\ }\textbf {\bibinfo {volume} {14}},\ \bibinfo
  {pages} {7035} (\bibinfo {year} {2023})}\BibitemShut {NoStop}%
\bibitem [{\citenamefont {Zhao}\ \emph {et~al.}(2026)\citenamefont {Zhao},
  \citenamefont {Yu}, \citenamefont {Ko{\v s}mrlj},\ and\ \citenamefont
  {Datta}}]{zhao2026dynamic}%
  \BibitemOpen
  \bibfield  {author} {\bibinfo {author} {\bibfnamefont {H.}~\bibnamefont
  {Zhao}}, \bibinfo {author} {\bibfnamefont {Q.}~\bibnamefont {Yu}}, \bibinfo
  {author} {\bibfnamefont {A.}~\bibnamefont {Ko{\v s}mrlj}},\ and\ \bibinfo
  {author} {\bibfnamefont {S.~S.}\ \bibnamefont {Datta}},\ }\bibfield  {title}
  {\bibinfo {title} {Dynamic instabilities and pattern formation in chemotactic
  active matter},\ }\Eprint {https://arxiv.org/abs/2603.02595}
  {arXiv:2603.02595}  (\bibinfo {year} {2026})\BibitemShut {NoStop}%
\bibitem [{\citenamefont {Liu}\ \emph {et~al.}(2024)\citenamefont {Liu},
  \citenamefont {Kailasham}, \citenamefont {Moerman}, \citenamefont {Khair},\
  and\ \citenamefont {Zarzar}}]{liu2024self}%
  \BibitemOpen
  \bibfield  {author} {\bibinfo {author} {\bibfnamefont {Y.}~\bibnamefont
  {Liu}}, \bibinfo {author} {\bibfnamefont {R.}~\bibnamefont {Kailasham}},
  \bibinfo {author} {\bibfnamefont {P.~G.}\ \bibnamefont {Moerman}}, \bibinfo
  {author} {\bibfnamefont {A.~S.}\ \bibnamefont {Khair}},\ and\ \bibinfo
  {author} {\bibfnamefont {L.~D.}\ \bibnamefont {Zarzar}},\ }\bibfield  {title}
  {\bibinfo {title} {Self-organized patterns in non-reciprocal active droplet
  systems},\ }\href {https://doi.org/10.1002/ange.202409382} {\bibfield
  {journal} {\bibinfo  {journal} {Angew. Chem., Int. Ed.}\ }\textbf {\bibinfo
  {volume} {63}},\ \bibinfo {pages} {e202409382} (\bibinfo {year}
  {2024})}\BibitemShut {NoStop}%
\bibitem [{\citenamefont {Long}\ \emph {et~al.}(2017)\citenamefont {Long},
  \citenamefont {Quaife}, \citenamefont {Salman},\ and\ \citenamefont
  {Oltvai}}]{long2017cell}%
  \BibitemOpen
  \bibfield  {author} {\bibinfo {author} {\bibfnamefont {Z.}~\bibnamefont
  {Long}}, \bibinfo {author} {\bibfnamefont {B.}~\bibnamefont {Quaife}},
  \bibinfo {author} {\bibfnamefont {H.}~\bibnamefont {Salman}},\ and\ \bibinfo
  {author} {\bibfnamefont {Z.~N.}\ \bibnamefont {Oltvai}},\ }\bibfield  {title}
  {\bibinfo {title} {Cell-cell communication enhances bacterial chemotaxis
  toward external attractants},\ }\href
  {https://doi.org/10.1038/s41598-017-13183-9} {\bibfield  {journal} {\bibinfo
  {journal} {Sci. Rep.}\ }\textbf {\bibinfo {volume} {7}},\ \bibinfo {pages}
  {12855} (\bibinfo {year} {2017})}\BibitemShut {NoStop}%
\bibitem [{\citenamefont {Saha}\ \emph {et~al.}(2026)\citenamefont {Saha},
  \citenamefont {Fancher},\ and\ \citenamefont {Mugler}}]{saha2026effects}%
  \BibitemOpen
  \bibfield  {author} {\bibinfo {author} {\bibfnamefont {S.}~\bibnamefont
  {Saha}}, \bibinfo {author} {\bibfnamefont {S.}~\bibnamefont {Fancher}},\ and\
  \bibinfo {author} {\bibfnamefont {A.}~\bibnamefont {Mugler}},\ }\bibfield
  {title} {\bibinfo {title} {Effects of cell-cell communication on bacterial
  chemotaxis},\ }\href {https://doi.org/10.1103/jgm1-2lcd} {\bibfield
  {journal} {\bibinfo  {journal} {Phys. Rev. E}\ }\textbf {\bibinfo {volume}
  {113}},\ \bibinfo {pages} {054415} (\bibinfo {year} {2026})}\BibitemShut
  {NoStop}%
\bibitem [{\citenamefont {Salman}\ \emph {et~al.}(2006)\citenamefont {Salman},
  \citenamefont {Zilman}, \citenamefont {Loverdo}, \citenamefont {Jeffroy},\
  and\ \citenamefont {Libchaber}}]{salman2006solitary}%
  \BibitemOpen
  \bibfield  {author} {\bibinfo {author} {\bibfnamefont {H.}~\bibnamefont
  {Salman}}, \bibinfo {author} {\bibfnamefont {A.}~\bibnamefont {Zilman}},
  \bibinfo {author} {\bibfnamefont {C.}~\bibnamefont {Loverdo}}, \bibinfo
  {author} {\bibfnamefont {M.}~\bibnamefont {Jeffroy}},\ and\ \bibinfo {author}
  {\bibfnamefont {A.}~\bibnamefont {Libchaber}},\ }\bibfield  {title} {\bibinfo
  {title} {Solitary modes of bacterial culture in a temperature gradient},\
  }\href {https://doi.org/10.1103/physrevlett.97.118101} {\bibfield  {journal}
  {\bibinfo  {journal} {Phys. Rev. Lett.}\ }\textbf {\bibinfo {volume} {97}},\
  \bibinfo {pages} {118101} (\bibinfo {year} {2006})}\BibitemShut {NoStop}%
\bibitem [{\citenamefont {Saragosti}\ \emph {et~al.}(2010)\citenamefont
  {Saragosti}, \citenamefont {Calvez}, \citenamefont {Bournaveas},
  \citenamefont {Buguin}, \citenamefont {Silberzan},\ and\ \citenamefont
  {Perthame}}]{saragosti2010mathematical}%
  \BibitemOpen
  \bibfield  {author} {\bibinfo {author} {\bibfnamefont {J.}~\bibnamefont
  {Saragosti}}, \bibinfo {author} {\bibfnamefont {V.}~\bibnamefont {Calvez}},
  \bibinfo {author} {\bibfnamefont {N.}~\bibnamefont {Bournaveas}}, \bibinfo
  {author} {\bibfnamefont {A.}~\bibnamefont {Buguin}}, \bibinfo {author}
  {\bibfnamefont {P.}~\bibnamefont {Silberzan}},\ and\ \bibinfo {author}
  {\bibfnamefont {B.}~\bibnamefont {Perthame}},\ }\bibfield  {title} {\bibinfo
  {title} {Mathematical description of bacterial traveling pulses},\ }\href
  {https://doi.org/10.1371/journal.pcbi.1000890} {\bibfield  {journal}
  {\bibinfo  {journal} {PLoS Comput. Biol.}\ }\textbf {\bibinfo {volume} {6}},\
  \bibinfo {pages} {e1000890} (\bibinfo {year} {2010})}\BibitemShut {NoStop}%
\bibitem [{\citenamefont {Young}\ \emph {et~al.}(2019)\citenamefont {Young},
  \citenamefont {Demir}, \citenamefont {Salman}, \citenamefont {Ermentrout},\
  and\ \citenamefont {Rubin}}]{young2019interactions}%
  \BibitemOpen
  \bibfield  {author} {\bibinfo {author} {\bibfnamefont {G.}~\bibnamefont
  {Young}}, \bibinfo {author} {\bibfnamefont {M.}~\bibnamefont {Demir}},
  \bibinfo {author} {\bibfnamefont {H.}~\bibnamefont {Salman}}, \bibinfo
  {author} {\bibfnamefont {G.~B.}\ \bibnamefont {Ermentrout}},\ and\ \bibinfo
  {author} {\bibfnamefont {J.~E.}\ \bibnamefont {Rubin}},\ }\bibfield  {title}
  {\bibinfo {title} {Interactions of solitary pulses of \textit{{E}. coli} in a
  one-dimensional nutrient gradient},\ }\href
  {https://doi.org/10.1016/j.physd.2019.02.007} {\bibfield  {journal} {\bibinfo
   {journal} {Physica D: Nonlinear Phenom.}\ }\textbf {\bibinfo {volume}
  {395}},\ \bibinfo {pages} {24} (\bibinfo {year} {2019})}\BibitemShut
  {NoStop}%
\bibitem [{\citenamefont {Perez~Ipi{\~n}a}\ and\ \citenamefont
  {Camley}(2026)}]{perez2026competing}%
  \BibitemOpen
  \bibfield  {author} {\bibinfo {author} {\bibfnamefont {E.}~\bibnamefont
  {Perez~Ipi{\~n}a}}\ and\ \bibinfo {author} {\bibfnamefont {B.~A.}\
  \bibnamefont {Camley}},\ }\bibfield  {title} {\bibinfo {title} {Competing
  chemical gradients change chemotactic dynamics and cell distribution},\
  }\href {https://doi.org/10.1103/z3zg-xrxx} {\bibfield  {journal} {\bibinfo
  {journal} {Phys. Rev. E}\ }\textbf {\bibinfo {volume} {113}},\ \bibinfo
  {pages} {034406} (\bibinfo {year} {2026})}\BibitemShut {NoStop}%
\bibitem [{\citenamefont {Curk}\ \emph {et~al.}(2013)\citenamefont {Curk},
  \citenamefont {Marenduzzo},\ and\ \citenamefont
  {Dobnikar}}]{curk2013chemotactic}%
  \BibitemOpen
  \bibfield  {author} {\bibinfo {author} {\bibfnamefont {T.}~\bibnamefont
  {Curk}}, \bibinfo {author} {\bibfnamefont {D.}~\bibnamefont {Marenduzzo}},\
  and\ \bibinfo {author} {\bibfnamefont {J.}~\bibnamefont {Dobnikar}},\
  }\bibfield  {title} {\bibinfo {title} {Chemotactic sensing towards ambient
  and secreted attractant drives collective behaviour of \textit{{E}. coli}},\
  }\href {https://doi.org/10.1371/journal.pone.0074878} {\bibfield  {journal}
  {\bibinfo  {journal} {PLoS One}\ }\textbf {\bibinfo {volume} {8}},\ \bibinfo
  {pages} {e74878} (\bibinfo {year} {2013})}\BibitemShut {NoStop}%
\bibitem [{not()}]{noteSI}%
  \BibitemOpen
  \bibinfo {note} {Supplemental material is provided at
  http://link.aps.org/supplemental/XXX}\BibitemShut {NoStop}%
\bibitem [{\citenamefont {Hillen}\ and\ \citenamefont
  {Painter}(2001)}]{hillen2001global}%
  \BibitemOpen
  \bibfield  {author} {\bibinfo {author} {\bibfnamefont {T.}~\bibnamefont
  {Hillen}}\ and\ \bibinfo {author} {\bibfnamefont {K.}~\bibnamefont
  {Painter}},\ }\bibfield  {title} {\bibinfo {title} {Global existence for a
  parabolic chemotaxis model with prevention of overcrowding},\ }\href
  {https://doi.org/10.1006/aama.2001.0721} {\bibfield  {journal} {\bibinfo
  {journal} {Adv. Appl. Math.}\ }\textbf {\bibinfo {volume} {26}},\ \bibinfo
  {pages} {280} (\bibinfo {year} {2001})}\BibitemShut {NoStop}%
\bibitem [{\citenamefont {Painter}\ \emph {et~al.}(2002)\citenamefont
  {Painter}, \citenamefont {Hillen} \emph {et~al.}}]{painter2002volume}%
  \BibitemOpen
  \bibfield  {author} {\bibinfo {author} {\bibfnamefont {K.~J.}\ \bibnamefont
  {Painter}}, \bibinfo {author} {\bibfnamefont {T.}~\bibnamefont {Hillen}},
  \emph {et~al.},\ }\bibfield  {title} {\bibinfo {title} {Volume-filling and
  quorum-sensing in models for chemosensitive movement},\ }\href@noop {}
  {\bibfield  {journal} {\bibinfo  {journal} {Can. Appl. Math. Quart}\ }\textbf
  {\bibinfo {volume} {10}},\ \bibinfo {pages} {501} (\bibinfo {year}
  {2002})}\BibitemShut {NoStop}%
\bibitem [{\citenamefont {Childress}(1984)}]{childress1984chemotactic}%
  \BibitemOpen
  \bibfield  {author} {\bibinfo {author} {\bibfnamefont {S.}~\bibnamefont
  {Childress}},\ }\bibinfo {title} {Chemotactic collapse in two dimensions},\
  in\ \href {https://doi.org/10.1007/978-3-642-45589-6_6} {\emph {\bibinfo
  {booktitle} {Modelling of Patterns in Space and Time: {{Proceedings}} of a
  Workshop Held by the Sonderforschungsbereich 123 at Heidelberg July 4--8,
  1983}}},\ \bibinfo {series} {Lecture Notes in Biomathematics}, Vol.~\bibinfo
  {volume} {55},\ \bibinfo {editor} {edited by\ \bibinfo {editor}
  {\bibfnamefont {W.}~\bibnamefont {J{\"a}ger}}\ and\ \bibinfo {editor}
  {\bibfnamefont {J.~D.}\ \bibnamefont {Murray}}}\ (\bibinfo  {publisher}
  {Springer},\ \bibinfo {year} {1984})\ pp.\ \bibinfo {pages}
  {61--66}\BibitemShut {NoStop}%
\bibitem [{\citenamefont {Herrero}\ and\ \citenamefont
  {Vel{\'a}zquez}(1996)}]{herrero1996chemotactic}%
  \BibitemOpen
  \bibfield  {author} {\bibinfo {author} {\bibfnamefont {M.~A.}\ \bibnamefont
  {Herrero}}\ and\ \bibinfo {author} {\bibfnamefont {J.~J.}\ \bibnamefont
  {Vel{\'a}zquez}},\ }\bibfield  {title} {\bibinfo {title} {Chemotactic
  collapse for the {{Keller-Segel}} model},\ }\href
  {https://doi.org/10.1007/s002850050049} {\bibfield  {journal} {\bibinfo
  {journal} {J. Math. Biol.}\ }\textbf {\bibinfo {volume} {35}},\ \bibinfo
  {pages} {177} (\bibinfo {year} {1996})}\BibitemShut {NoStop}%
\bibitem [{\citenamefont {Childress}\ and\ \citenamefont
  {Percus}(1981)}]{childress1981nonlinear}%
  \BibitemOpen
  \bibfield  {author} {\bibinfo {author} {\bibfnamefont {S.}~\bibnamefont
  {Childress}}\ and\ \bibinfo {author} {\bibfnamefont {J.~K.}\ \bibnamefont
  {Percus}},\ }\bibfield  {title} {\bibinfo {title} {Nonlinear aspects of
  chemotaxis},\ }\href {https://doi.org/10.1016/0025-5564(81)90055-9}
  {\bibfield  {journal} {\bibinfo  {journal} {Math. Biosci.}\ }\textbf
  {\bibinfo {volume} {56}},\ \bibinfo {pages} {217} (\bibinfo {year}
  {1981})}\BibitemShut {NoStop}%
\bibitem [{\citenamefont {Brauns}\ and\ \citenamefont
  {Marchetti}(2024)}]{brauns2024nonreciprocal}%
  \BibitemOpen
  \bibfield  {author} {\bibinfo {author} {\bibfnamefont {F.}~\bibnamefont
  {Brauns}}\ and\ \bibinfo {author} {\bibfnamefont {M.~C.}\ \bibnamefont
  {Marchetti}},\ }\bibfield  {title} {\bibinfo {title} {Nonreciprocal pattern
  formation of conserved fields},\ }\href
  {https://doi.org/10.1103/physrevx.14.021014} {\bibfield  {journal} {\bibinfo
  {journal} {Phys. Rev. X}\ }\textbf {\bibinfo {volume} {14}},\ \bibinfo
  {pages} {021014} (\bibinfo {year} {2024})}\BibitemShut {NoStop}%
\bibitem [{\citenamefont {Winkler}\ \emph {et~al.}(2026)\citenamefont
  {Winkler}, \citenamefont {Alonso},\ and\ \citenamefont
  {B{\"a}r}}]{winkler2026simple}%
  \BibitemOpen
  \bibfield  {author} {\bibinfo {author} {\bibfnamefont {B.}~\bibnamefont
  {Winkler}}, \bibinfo {author} {\bibfnamefont {S.}~\bibnamefont {Alonso}},\
  and\ \bibinfo {author} {\bibfnamefont {M.}~\bibnamefont {B{\"a}r}},\
  }\bibfield  {title} {\bibinfo {title} {A simple model for conserved
  intracellular dynamics exhibits multiscale pattern formation, traveling
  protein domains and arrested coarsening of lipid domains},\ }\Eprint
  {https://arxiv.org/abs/2603.02595} {arXiv:2603.02595}  (\bibinfo {year}
  {2026})\BibitemShut {NoStop}%
\bibitem [{\citenamefont {Brauns}\ \emph {et~al.}(2021)\citenamefont {Brauns},
  \citenamefont {Weyer}, \citenamefont {Halatek}, \citenamefont {Yoon},\ and\
  \citenamefont {Frey}}]{brauns2021wavelength}%
  \BibitemOpen
  \bibfield  {author} {\bibinfo {author} {\bibfnamefont {F.}~\bibnamefont
  {Brauns}}, \bibinfo {author} {\bibfnamefont {H.}~\bibnamefont {Weyer}},
  \bibinfo {author} {\bibfnamefont {J.}~\bibnamefont {Halatek}}, \bibinfo
  {author} {\bibfnamefont {J.}~\bibnamefont {Yoon}},\ and\ \bibinfo {author}
  {\bibfnamefont {E.}~\bibnamefont {Frey}},\ }\bibfield  {title} {\bibinfo
  {title} {Wavelength selection by interrupted coarsening in reaction-diffusion
  systems},\ }\href {https://doi.org/10.1103/physrevlett.126.104101} {\bibfield
   {journal} {\bibinfo  {journal} {Phys. Rev. Lett.}\ }\textbf {\bibinfo
  {volume} {126}},\ \bibinfo {pages} {104101} (\bibinfo {year}
  {2021})}\BibitemShut {NoStop}%
\bibitem [{\citenamefont {Weyer}\ \emph {et~al.}(2023)\citenamefont {Weyer},
  \citenamefont {Brauns},\ and\ \citenamefont {Frey}}]{weyer2023coarsening}%
  \BibitemOpen
  \bibfield  {author} {\bibinfo {author} {\bibfnamefont {H.}~\bibnamefont
  {Weyer}}, \bibinfo {author} {\bibfnamefont {F.}~\bibnamefont {Brauns}},\ and\
  \bibinfo {author} {\bibfnamefont {E.}~\bibnamefont {Frey}},\ }\bibfield
  {title} {\bibinfo {title} {Coarsening and wavelength selection far from
  equilibrium: {{A}} unifying framework based on singular perturbation
  theory},\ }\href {https://doi.org/10.1103/physreve.108.064202} {\bibfield
  {journal} {\bibinfo  {journal} {Phys. Rev. E}\ }\textbf {\bibinfo {volume}
  {108}},\ \bibinfo {pages} {064202} (\bibinfo {year} {2023})}\BibitemShut
  {NoStop}%
\bibitem [{\citenamefont {Frey}\ and\ \citenamefont
  {Weyer}(2026)}]{frey2026pattern}%
  \BibitemOpen
  \bibfield  {author} {\bibinfo {author} {\bibfnamefont {E.}~\bibnamefont
  {Frey}}\ and\ \bibinfo {author} {\bibfnamefont {H.}~\bibnamefont {Weyer}},\
  }\bibfield  {title} {\bibinfo {title} {Pattern formation beyond {{Turing}}:
  {{Physical}} principles of mass-conserving reaction-diffusion systems},\
  }\href {https://doi.org/10.1146/annurev-biophys-030822-031638} {\bibfield
  {journal} {\bibinfo  {journal} {Ann. Rev. Biophys.}\ }\textbf {\bibinfo
  {volume} {55}},\ \bibinfo {pages} {030822} (\bibinfo {year}
  {2026})}\BibitemShut {NoStop}%
\bibitem [{\citenamefont {Weyer}\ \emph
  {et~al.}(2025{\natexlab{a}})\citenamefont {Weyer}, \citenamefont
  {Muramatsu},\ and\ \citenamefont {Frey}}]{weyer2025coarsening}%
  \BibitemOpen
  \bibfield  {author} {\bibinfo {author} {\bibfnamefont {H.}~\bibnamefont
  {Weyer}}, \bibinfo {author} {\bibfnamefont {D.}~\bibnamefont {Muramatsu}},\
  and\ \bibinfo {author} {\bibfnamefont {E.}~\bibnamefont {Frey}},\ }\bibfield
  {title} {\bibinfo {title} {Coarsening dynamics of chemotactic aggregates},\
  }\href {https://doi.org/10.1103/c9px-mdhs} {\bibfield  {journal} {\bibinfo
  {journal} {Phys. Rev. E}\ }\textbf {\bibinfo {volume} {112}},\ \bibinfo
  {pages} {054406} (\bibinfo {year} {2025}{\natexlab{a}})}\BibitemShut
  {NoStop}%
\bibitem [{\citenamefont {Weyer}\ \emph
  {et~al.}(2025{\natexlab{b}})\citenamefont {Weyer}, \citenamefont
  {Muramatsu},\ and\ \citenamefont {Frey}}]{weyer2025chemotaxis}%
  \BibitemOpen
  \bibfield  {author} {\bibinfo {author} {\bibfnamefont {H.}~\bibnamefont
  {Weyer}}, \bibinfo {author} {\bibfnamefont {D.}~\bibnamefont {Muramatsu}},\
  and\ \bibinfo {author} {\bibfnamefont {E.}~\bibnamefont {Frey}},\ }\bibfield
  {title} {\bibinfo {title} {Chemotaxis-induced phase separation},\ }\href
  {https://doi.org/10.1103/2933-45qc} {\bibfield  {journal} {\bibinfo
  {journal} {Phys. Rev. Lett.}\ }\textbf {\bibinfo {volume} {135}},\ \bibinfo
  {pages} {208402} (\bibinfo {year} {2025}{\natexlab{b}})}\BibitemShut
  {NoStop}%
\bibitem [{tra()}]{transition_line}%
  \BibitemOpen
  \bibinfo {note} {Since the full boundary-value simulations start only from a
  perturbed homogeneous state and develop disordered aggregates with a
  distribution of spacings and masses, this calculation is not intended as an
  exact stability boundary. Rather, $\Lambda$ is interpreted as a
  representative local inter-aggregate spacing, and the resulting transition
  line provides a phenomenological mechanistic estimate of which coarsening
  pathway is expected to dominate statistically in the highlighted
  region.}\BibitemShut {Stop}%
\bibitem [{\citenamefont {McDougald}\ \emph {et~al.}(2012)\citenamefont
  {McDougald}, \citenamefont {Rice}, \citenamefont {Barraud}, \citenamefont
  {Steinberg},\ and\ \citenamefont {Kjelleberg}}]{mcdougald2012should}%
  \BibitemOpen
  \bibfield  {author} {\bibinfo {author} {\bibfnamefont {D.}~\bibnamefont
  {McDougald}}, \bibinfo {author} {\bibfnamefont {S.~A.}\ \bibnamefont {Rice}},
  \bibinfo {author} {\bibfnamefont {N.}~\bibnamefont {Barraud}}, \bibinfo
  {author} {\bibfnamefont {P.~D.}\ \bibnamefont {Steinberg}},\ and\ \bibinfo
  {author} {\bibfnamefont {S.}~\bibnamefont {Kjelleberg}},\ }\bibfield  {title}
  {\bibinfo {title} {Should we stay or should we go: mechanisms and ecological
  consequences for biofilm dispersal},\ }\href
  {https://doi.org/10.1038/nrmicro2695} {\bibfield  {journal} {\bibinfo
  {journal} {Nat. Rev. Microbiol.}\ }\textbf {\bibinfo {volume} {10}},\
  \bibinfo {pages} {39} (\bibinfo {year} {2012})}\BibitemShut {NoStop}%
\bibitem [{\citenamefont {Laganenka}\ \emph {et~al.}(2016)\citenamefont
  {Laganenka}, \citenamefont {Colin},\ and\ \citenamefont
  {Sourjik}}]{laganenka2016}%
  \BibitemOpen
  \bibfield  {author} {\bibinfo {author} {\bibfnamefont {L.}~\bibnamefont
  {Laganenka}}, \bibinfo {author} {\bibfnamefont {R.}~\bibnamefont {Colin}},\
  and\ \bibinfo {author} {\bibfnamefont {V.}~\bibnamefont {Sourjik}},\
  }\bibfield  {title} {\bibinfo {title} {Chemotaxis towards autoinducer 2
  mediates autoaggregation in {{\textit{Escherichia coli}}}},\ }\href
  {https://doi.org/10.1038/ncomms12984} {\bibfield  {journal} {\bibinfo
  {journal} {Nat. Commun.}\ }\textbf {\bibinfo {volume} {7}},\ \bibinfo {pages}
  {12984} (\bibinfo {year} {2016})}\BibitemShut {NoStop}%
\bibitem [{\citenamefont {Lushi}\ \emph {et~al.}(2018)\citenamefont {Lushi},
  \citenamefont {Goldstein},\ and\ \citenamefont
  {Shelley}}]{lushi2018nonlinear}%
  \BibitemOpen
  \bibfield  {author} {\bibinfo {author} {\bibfnamefont {E.}~\bibnamefont
  {Lushi}}, \bibinfo {author} {\bibfnamefont {R.~E.}\ \bibnamefont
  {Goldstein}},\ and\ \bibinfo {author} {\bibfnamefont {M.~J.}\ \bibnamefont
  {Shelley}},\ }\bibfield  {title} {\bibinfo {title} {Nonlinear concentration
  patterns and bands in autochemotactic suspensions},\ }\href
  {https://doi.org/10.1103/physreve.98.052411} {\bibfield  {journal} {\bibinfo
  {journal} {Phys. Rev. E}\ }\textbf {\bibinfo {volume} {98}},\ \bibinfo
  {pages} {052411} (\bibinfo {year} {2018})}\BibitemShut {NoStop}%
\bibitem [{\citenamefont {Theurkauff}\ \emph {et~al.}(2012)\citenamefont
  {Theurkauff}, \citenamefont {{Cottin-Bizonne}}, \citenamefont {Palacci},
  \citenamefont {Ybert},\ and\ \citenamefont
  {Bocquet}}]{theurkauff2012dynamic}%
  \BibitemOpen
  \bibfield  {author} {\bibinfo {author} {\bibfnamefont {I.}~\bibnamefont
  {Theurkauff}}, \bibinfo {author} {\bibfnamefont {C.}~\bibnamefont
  {{Cottin-Bizonne}}}, \bibinfo {author} {\bibfnamefont {J.}~\bibnamefont
  {Palacci}}, \bibinfo {author} {\bibfnamefont {C.}~\bibnamefont {Ybert}},\
  and\ \bibinfo {author} {\bibfnamefont {L.}~\bibnamefont {Bocquet}},\
  }\bibfield  {title} {\bibinfo {title} {Dynamic clustering in active colloidal
  suspensions with chemical signaling},\ }\href
  {https://doi.org/10.1103/physrevlett.108.268303} {\bibfield  {journal}
  {\bibinfo  {journal} {Phys. Rev. Lett.}\ }\textbf {\bibinfo {volume} {108}},\
  \bibinfo {pages} {268303} (\bibinfo {year} {2012})}\BibitemShut {NoStop}%
\bibitem [{\citenamefont {Pohl}\ and\ \citenamefont
  {Stark}(2014)}]{pohl2014dynamic}%
  \BibitemOpen
  \bibfield  {author} {\bibinfo {author} {\bibfnamefont {O.}~\bibnamefont
  {Pohl}}\ and\ \bibinfo {author} {\bibfnamefont {H.}~\bibnamefont {Stark}},\
  }\bibfield  {title} {\bibinfo {title} {Dynamic clustering and chemotactic
  collapse of self-phoretic active particles},\ }\href
  {https://doi.org/10.1103/physrevlett.112.238303} {\bibfield  {journal}
  {\bibinfo  {journal} {Phys. Rev. Lett.}\ }\textbf {\bibinfo {volume} {112}},\
  \bibinfo {pages} {238303} (\bibinfo {year} {2014})}\BibitemShut {NoStop}%
\bibitem [{\citenamefont {Liebchen}\ \emph {et~al.}(2017)\citenamefont
  {Liebchen}, \citenamefont {Marenduzzo},\ and\ \citenamefont
  {Cates}}]{liebchen2017phoretic}%
  \BibitemOpen
  \bibfield  {author} {\bibinfo {author} {\bibfnamefont {B.}~\bibnamefont
  {Liebchen}}, \bibinfo {author} {\bibfnamefont {D.}~\bibnamefont
  {Marenduzzo}},\ and\ \bibinfo {author} {\bibfnamefont {M.~E.}\ \bibnamefont
  {Cates}},\ }\bibfield  {title} {\bibinfo {title} {Phoretic interactions
  generically induce dynamic clusters and wave patterns in active colloids},\
  }\href {https://doi.org/10.1103/physrevlett.118.268001} {\bibfield  {journal}
  {\bibinfo  {journal} {Phys. Rev. Lett.}\ }\textbf {\bibinfo {volume} {118}},\
  \bibinfo {pages} {268001} (\bibinfo {year} {2017})}\BibitemShut {NoStop}%
\bibitem [{\citenamefont {Liebchen}\ and\ \citenamefont
  {L{\"o}wen}(2018)}]{liebchen2018synthetic}%
  \BibitemOpen
  \bibfield  {author} {\bibinfo {author} {\bibfnamefont {B.}~\bibnamefont
  {Liebchen}}\ and\ \bibinfo {author} {\bibfnamefont {H.}~\bibnamefont
  {L{\"o}wen}},\ }\bibfield  {title} {\bibinfo {title} {Synthetic chemotaxis
  and collective behavior in active matter},\ }\href
  {https://doi.org/10.1021/acs.accounts.8b00215} {\bibfield  {journal}
  {\bibinfo  {journal} {Acc. Chem. Res.}\ }\textbf {\bibinfo {volume} {51}},\
  \bibinfo {pages} {2982} (\bibinfo {year} {2018})}\BibitemShut {NoStop}%
\bibitem [{\citenamefont {Stark}(2018)}]{stark2018artificial}%
  \BibitemOpen
  \bibfield  {author} {\bibinfo {author} {\bibfnamefont {H.}~\bibnamefont
  {Stark}},\ }\bibfield  {title} {\bibinfo {title} {Artificial chemotaxis of
  self-phoretic active colloids: {C}ollective behavior},\ }\href
  {https://doi.org/10.1021/acs.accounts.8b00259} {\bibfield  {journal}
  {\bibinfo  {journal} {Accounts Chem. Res.}\ }\textbf {\bibinfo {volume}
  {51}},\ \bibinfo {pages} {2681} (\bibinfo {year} {2018})}\BibitemShut
  {NoStop}%
\bibitem [{\citenamefont {{Agudo-Canalejo}}\ and\ \citenamefont
  {Golestanian}(2019)}]{agudo2019active}%
  \BibitemOpen
  \bibfield  {author} {\bibinfo {author} {\bibfnamefont {J.}~\bibnamefont
  {{Agudo-Canalejo}}}\ and\ \bibinfo {author} {\bibfnamefont {R.}~\bibnamefont
  {Golestanian}},\ }\bibfield  {title} {\bibinfo {title} {Active phase
  separation in mixtures of chemically interacting particles},\ }\href
  {https://doi.org/10.1103/physrevlett.123.018101} {\bibfield  {journal}
  {\bibinfo  {journal} {Phys. Rev. Lett.}\ }\textbf {\bibinfo {volume} {123}},\
  \bibinfo {pages} {018101} (\bibinfo {year} {2019})}\BibitemShut {NoStop}%
\bibitem [{\citenamefont {Cates}\ and\ \citenamefont
  {Nardini}(2025)}]{cates2025active}%
  \BibitemOpen
  \bibfield  {author} {\bibinfo {author} {\bibfnamefont {M.~E.}\ \bibnamefont
  {Cates}}\ and\ \bibinfo {author} {\bibfnamefont {C.}~\bibnamefont
  {Nardini}},\ }\bibfield  {title} {\bibinfo {title} {Active phase separation:
  {N}ew phenomenology from non-equilibrium physics},\ }\href
  {https://doi.org/10.1088/1361-6633/add278} {\bibfield  {journal} {\bibinfo
  {journal} {Rep. Prog. Phys.}\ }\textbf {\bibinfo {volume} {88}},\ \bibinfo
  {pages} {056601} (\bibinfo {year} {2025})}\BibitemShut {NoStop}%
\bibitem [{\citenamefont {Osmanovi{\'c}}\ and\ \citenamefont
  {Franco}(2025)}]{osmanovic2025complex}%
  \BibitemOpen
  \bibfield  {author} {\bibinfo {author} {\bibfnamefont {D.}~\bibnamefont
  {Osmanovi{\'c}}}\ and\ \bibinfo {author} {\bibfnamefont {E.}~\bibnamefont
  {Franco}},\ }\bibfield  {title} {\bibinfo {title} {Complex dynamics in
  reaction-phase separation systems},\ }\href
  {https://doi.org/10.1103/physreve.111.025404} {\bibfield  {journal} {\bibinfo
   {journal} {Phys. Rev. E}\ }\textbf {\bibinfo {volume} {111}},\ \bibinfo
  {pages} {025404} (\bibinfo {year} {2025})}\BibitemShut {NoStop}%
\bibitem [{\citenamefont {Willems}\ \emph {et~al.}(2025)\citenamefont
  {Willems}, \citenamefont {Baron}, \citenamefont {Fernandez-Matoz},
  \citenamefont {Wolfisberg}, \citenamefont {Baret}, \citenamefont {Dufresne},\
  and\ \citenamefont {Alvarez}}]{willems2025run}%
  \BibitemOpen
  \bibfield  {author} {\bibinfo {author} {\bibfnamefont {V.}~\bibnamefont
  {Willems}}, \bibinfo {author} {\bibfnamefont {A.}~\bibnamefont {Baron}},
  \bibinfo {author} {\bibfnamefont {D.}~\bibnamefont {Fernandez-Matoz}},
  \bibinfo {author} {\bibfnamefont {G.}~\bibnamefont {Wolfisberg}}, \bibinfo
  {author} {\bibfnamefont {J.-C.}\ \bibnamefont {Baret}}, \bibinfo {author}
  {\bibfnamefont {E.}~\bibnamefont {Dufresne}},\ and\ \bibinfo {author}
  {\bibfnamefont {L.}~\bibnamefont {Alvarez}},\ }\bibfield  {title} {\bibinfo
  {title} {Run-and-tumble dynamics of active giant vesicles},\ }\href
  {https://doi.org/10.1039/d5sm00309a} {\bibfield  {journal} {\bibinfo
  {journal} {Soft Matter}\ }\textbf {\bibinfo {volume} {21}},\ \bibinfo {pages}
  {6175} (\bibinfo {year} {2025})}\BibitemShut {NoStop}%
\bibitem [{\citenamefont {Oza}\ \emph {et~al.}(2026)\citenamefont {Oza},
  \citenamefont {Kanso},\ and\ \citenamefont {Shelley}}]{oza2026traveling}%
  \BibitemOpen
  \bibfield  {author} {\bibinfo {author} {\bibfnamefont {A.~U.}\ \bibnamefont
  {Oza}}, \bibinfo {author} {\bibfnamefont {E.}~\bibnamefont {Kanso}},\ and\
  \bibinfo {author} {\bibfnamefont {M.~J.}\ \bibnamefont {Shelley}},\
  }\bibfield  {title} {\bibinfo {title} {Traveling waves in a continuum model
  of schooling swimmers},\ }\href {https://doi.org/10.1073/pnas.2521850123}
  {\bibfield  {journal} {\bibinfo  {journal} {Proc. Natl. Acad. Sci. USA}\
  }\textbf {\bibinfo {volume} {123}},\ \bibinfo {pages} {e2521850123} (\bibinfo
  {year} {2026})}\BibitemShut {NoStop}%
\bibitem [{\citenamefont {Borisova}\ \emph {et~al.}(2025)\citenamefont
  {Borisova}, \citenamefont {Paunova-Krasteva}, \citenamefont {Strateva},\ and\
  \citenamefont {Stoitsova}}]{borisova2025biofilm}%
  \BibitemOpen
  \bibfield  {author} {\bibinfo {author} {\bibfnamefont {D.}~\bibnamefont
  {Borisova}}, \bibinfo {author} {\bibfnamefont {T.}~\bibnamefont
  {Paunova-Krasteva}}, \bibinfo {author} {\bibfnamefont {T.}~\bibnamefont
  {Strateva}},\ and\ \bibinfo {author} {\bibfnamefont {S.}~\bibnamefont
  {Stoitsova}},\ }\bibfield  {title} {\bibinfo {title} {Biofilm formation of
  {{\textit{{P}seudomonas aeruginosa}}} in cystic fibrosis: Mechanisms of
  persistence, adaptation, and pathogenesis},\ }\href
  {https://doi.org/10.3390/microorganisms13071527} {\bibfield  {journal}
  {\bibinfo  {journal} {Microorganisms}\ }\textbf {\bibinfo {volume} {13}},\
  \bibinfo {pages} {1527} (\bibinfo {year} {2025})}\BibitemShut {NoStop}%
\bibitem [{\citenamefont {Jin}\ \emph {et~al.}(2017)\citenamefont {Jin},
  \citenamefont {Kr{\"u}ger},\ and\ \citenamefont {Maass}}]{jin2017chemotaxis}%
  \BibitemOpen
  \bibfield  {author} {\bibinfo {author} {\bibfnamefont {C.}~\bibnamefont
  {Jin}}, \bibinfo {author} {\bibfnamefont {C.}~\bibnamefont {Kr{\"u}ger}},\
  and\ \bibinfo {author} {\bibfnamefont {C.~C.}\ \bibnamefont {Maass}},\
  }\bibfield  {title} {\bibinfo {title} {Chemotaxis and autochemotaxis of
  self-propelling droplet swimmers},\ }\href
  {https://doi.org/10.1073/pnas.1619783114} {\bibfield  {journal} {\bibinfo
  {journal} {Proc. Natl. Acad. Sci. USA}\ }\textbf {\bibinfo {volume} {114}},\
  \bibinfo {pages} {5089} (\bibinfo {year} {2017})}\BibitemShut {NoStop}%
\bibitem [{\citenamefont {G{\'o}mez-Nava}\ \emph {et~al.}(2018)\citenamefont
  {G{\'o}mez-Nava}, \citenamefont {Gro{\ss{}}mann},\ and\ \citenamefont
  {Peruani}}]{nava2018markovian}%
  \BibitemOpen
  \bibfield  {author} {\bibinfo {author} {\bibfnamefont {L.}~\bibnamefont
  {G{\'o}mez-Nava}}, \bibinfo {author} {\bibfnamefont {R.}~\bibnamefont
  {Gro{\ss{}}mann}},\ and\ \bibinfo {author} {\bibfnamefont {F.}~\bibnamefont
  {Peruani}},\ }\bibfield  {title} {\bibinfo {title} {Markovian robots: Minimal
  navigation strategies for active particles},\ }\href
  {https://doi.org/10.1103/PhysRevE.97.042604} {\bibfield  {journal} {\bibinfo
  {journal} {Phys. Rev. E}\ }\textbf {\bibinfo {volume} {97}},\ \bibinfo
  {pages} {042604} (\bibinfo {year} {2018})}\BibitemShut {NoStop}%
\bibitem [{\citenamefont {Peng}\ and\ \citenamefont
  {Kapral}(2024)}]{peng2024self}%
  \BibitemOpen
  \bibfield  {author} {\bibinfo {author} {\bibfnamefont {Z.}~\bibnamefont
  {Peng}}\ and\ \bibinfo {author} {\bibfnamefont {R.}~\bibnamefont {Kapral}},\
  }\bibfield  {title} {\bibinfo {title} {Self-organization of active colloids
  mediated by chemical interactions},\ }\href
  {https://doi.org/10.1039/d3sm01272g} {\bibfield  {journal} {\bibinfo
  {journal} {Soft Matter}\ }\textbf {\bibinfo {volume} {20}},\ \bibinfo {pages}
  {1100} (\bibinfo {year} {2024})}\BibitemShut {NoStop}%
\bibitem [{\citenamefont {Xiong}\ \emph {et~al.}(2025)\citenamefont {Xiong},
  \citenamefont {Liu}, \citenamefont {Wang}, \citenamefont {Ong},\ and\
  \citenamefont {Zhu}}]{xiong2025chemotactic}%
  \BibitemOpen
  \bibfield  {author} {\bibinfo {author} {\bibfnamefont {T.}~\bibnamefont
  {Xiong}}, \bibinfo {author} {\bibfnamefont {Z.}~\bibnamefont {Liu}}, \bibinfo
  {author} {\bibfnamefont {Y.}~\bibnamefont {Wang}}, \bibinfo {author}
  {\bibfnamefont {C.~J.}\ \bibnamefont {Ong}},\ and\ \bibinfo {author}
  {\bibfnamefont {L.}~\bibnamefont {Zhu}},\ }\bibfield  {title} {\bibinfo
  {title} {Chemotactic navigation in robotic swimmers via reset-free
  hierarchical reinforcement learning},\ }\href
  {https://doi.org/10.1038/s41467-025-60646-z} {\bibfield  {journal} {\bibinfo
  {journal} {Nat. Commun.}\ }\textbf {\bibinfo {volume} {16}},\ \bibinfo
  {pages} {5441} (\bibinfo {year} {2025})}\BibitemShut {NoStop}%
\bibitem [{\citenamefont {Raman}\ \emph {et~al.}(2026)\citenamefont {Raman},
  \citenamefont {Shivhare}, \citenamefont {Kumar}, \citenamefont {Penukonda},
  \citenamefont {Kumar}, \citenamefont {Singh}, \citenamefont {Choudhary},\
  and\ \citenamefont {Mangal}}]{raman2026clustering}%
  \BibitemOpen
  \bibfield  {author} {\bibinfo {author} {\bibfnamefont {H.}~\bibnamefont
  {Raman}}, \bibinfo {author} {\bibfnamefont {A.}~\bibnamefont {Shivhare}},
  \bibinfo {author} {\bibfnamefont {A.}~\bibnamefont {Kumar}}, \bibinfo
  {author} {\bibfnamefont {M.}~\bibnamefont {Penukonda}}, \bibinfo {author}
  {\bibfnamefont {P.}~\bibnamefont {Kumar}}, \bibinfo {author} {\bibfnamefont
  {K.}~\bibnamefont {Singh}}, \bibinfo {author} {\bibfnamefont
  {A.}~\bibnamefont {Choudhary}},\ and\ \bibinfo {author} {\bibfnamefont
  {R.}~\bibnamefont {Mangal}},\ }\bibfield  {title} {\bibinfo {title}
  {Clustering dynamics of {{SiO2}}-{{Pt}} active {{Janus}} colloids},\ }\href
  {https://doi.org/10.1063/5.0321353} {\bibfield  {journal} {\bibinfo
  {journal} {Phys. Fluids}\ }\textbf {\bibinfo {volume} {38}},\ \bibinfo
  {pages} {062023} (\bibinfo {year} {2026})}\BibitemShut {NoStop}%
\bibitem [{\citenamefont {Wang}\ \emph {et~al.}(2026)\citenamefont {Wang},
  \citenamefont {Zhang}, \citenamefont {Chaudhari}, \citenamefont {Shim},
  \citenamefont {Severn}, \citenamefont {Zheng}, \citenamefont {Liang},
  \citenamefont {Ji}, \citenamefont {Pelster}, \citenamefont {Griniasty} \emph
  {et~al.}}]{wang2026microscopic}%
  \BibitemOpen
  \bibfield  {author} {\bibinfo {author} {\bibfnamefont {W.}~\bibnamefont
  {Wang}}, \bibinfo {author} {\bibfnamefont {J.}~\bibnamefont {Zhang}},
  \bibinfo {author} {\bibfnamefont {P.}~\bibnamefont {Chaudhari}}, \bibinfo
  {author} {\bibfnamefont {K.}~\bibnamefont {Shim}}, \bibinfo {author}
  {\bibfnamefont {J.}~\bibnamefont {Severn}}, \bibinfo {author} {\bibfnamefont
  {C.}~\bibnamefont {Zheng}}, \bibinfo {author} {\bibfnamefont
  {Z.}~\bibnamefont {Liang}}, \bibinfo {author} {\bibfnamefont
  {Y.}~\bibnamefont {Ji}}, \bibinfo {author} {\bibfnamefont {J.}~\bibnamefont
  {Pelster}}, \bibinfo {author} {\bibfnamefont {I.}~\bibnamefont {Griniasty}},
  \emph {et~al.},\ }\bibfield  {title} {\bibinfo {title} {Microscopic robots
  that sense and reshape their environment},\ }\href
  {https://doi.org/10.1038/s41928-026-01709-x} {\bibfield  {journal} {\bibinfo
  {journal} {Nat. Electron.}\ } (\bibinfo {year} {2026})}\BibitemShut {NoStop}%
\bibitem [{\citenamefont {O'Loughlin}\ \emph {et~al.}(2013)\citenamefont
  {O'Loughlin}, \citenamefont {Miller}, \citenamefont {Siryaporn},
  \citenamefont {Drescher}, \citenamefont {Semmelhack},\ and\ \citenamefont
  {Bassler}}]{oloughlin2013quorum}%
  \BibitemOpen
  \bibfield  {author} {\bibinfo {author} {\bibfnamefont {C.~T.}\ \bibnamefont
  {O'Loughlin}}, \bibinfo {author} {\bibfnamefont {L.~C.}\ \bibnamefont
  {Miller}}, \bibinfo {author} {\bibfnamefont {A.}~\bibnamefont {Siryaporn}},
  \bibinfo {author} {\bibfnamefont {K.}~\bibnamefont {Drescher}}, \bibinfo
  {author} {\bibfnamefont {M.~F.}\ \bibnamefont {Semmelhack}},\ and\ \bibinfo
  {author} {\bibfnamefont {B.~L.}\ \bibnamefont {Bassler}},\ }\bibfield
  {title} {\bibinfo {title} {A quorum-sensing inhibitor blocks
  {{\textit{{P}seudomonas aeruginosa}}} virulence and biofilm formation},\
  }\href {https://doi.org/10.1073/pnas.1316981110} {\bibfield  {journal}
  {\bibinfo  {journal} {Proc. Natl. Acad. Sci. USA}\ }\textbf {\bibinfo
  {volume} {110}},\ \bibinfo {pages} {17981} (\bibinfo {year}
  {2013})}\BibitemShut {NoStop}%
\bibitem [{\citenamefont {Cates}\ \emph {et~al.}(2010)\citenamefont {Cates},
  \citenamefont {Marenduzzo}, \citenamefont {Pagonabarraga},\ and\
  \citenamefont {Tailleur}}]{cates2010arrested}%
  \BibitemOpen
  \bibfield  {author} {\bibinfo {author} {\bibfnamefont {M.~E.}\ \bibnamefont
  {Cates}}, \bibinfo {author} {\bibfnamefont {D.}~\bibnamefont {Marenduzzo}},
  \bibinfo {author} {\bibfnamefont {I.}~\bibnamefont {Pagonabarraga}},\ and\
  \bibinfo {author} {\bibfnamefont {J.}~\bibnamefont {Tailleur}},\ }\bibfield
  {title} {\bibinfo {title} {Arrested phase separation in reproducing bacteria
  creates a generic route to pattern formation},\ }\href
  {https://doi.org/10.1073/pnas.1001994107} {\bibfield  {journal} {\bibinfo
  {journal} {Proc. Natl. Acad. Sci. USA}\ }\textbf {\bibinfo {volume} {107}},\
  \bibinfo {pages} {11715} (\bibinfo {year} {2010})}\BibitemShut {NoStop}%
\bibitem [{\citenamefont {Centler}\ \emph {et~al.}(2011)\citenamefont
  {Centler}, \citenamefont {Fetzer},\ and\ \citenamefont
  {Thullner}}]{centler2011modeling}%
  \BibitemOpen
  \bibfield  {author} {\bibinfo {author} {\bibfnamefont {F.}~\bibnamefont
  {Centler}}, \bibinfo {author} {\bibfnamefont {I.}~\bibnamefont {Fetzer}},\
  and\ \bibinfo {author} {\bibfnamefont {M.}~\bibnamefont {Thullner}},\
  }\bibfield  {title} {\bibinfo {title} {Modeling population patterns of
  chemotactic bacteria in homogeneous porous media},\ }\href
  {https://doi.org/10.1016/j.jtbi.2011.07.024} {\bibfield  {journal} {\bibinfo
  {journal} {J. Theor. Biol.}\ }\textbf {\bibinfo {volume} {287}},\ \bibinfo
  {pages} {82} (\bibinfo {year} {2011})}\BibitemShut {NoStop}%
\bibitem [{\citenamefont {Gelimson}\ and\ \citenamefont
  {Golestanian}(2015)}]{gelimson2015collective}%
  \BibitemOpen
  \bibfield  {author} {\bibinfo {author} {\bibfnamefont {A.}~\bibnamefont
  {Gelimson}}\ and\ \bibinfo {author} {\bibfnamefont {R.}~\bibnamefont
  {Golestanian}},\ }\bibfield  {title} {\bibinfo {title} {Collective dynamics
  of dividing chemotactic cells},\ }\href
  {https://doi.org/10.1103/physrevlett.114.028101} {\bibfield  {journal}
  {\bibinfo  {journal} {Phys. Rev. Lett.}\ }\textbf {\bibinfo {volume} {114}},\
  \bibinfo {pages} {028101} (\bibinfo {year} {2015})}\BibitemShut {NoStop}%
\bibitem [{\citenamefont {Mukherjee}\ and\ \citenamefont
  {Ghosh}(2018)}]{mukherjee2018growth}%
  \BibitemOpen
  \bibfield  {author} {\bibinfo {author} {\bibfnamefont {M.}~\bibnamefont
  {Mukherjee}}\ and\ \bibinfo {author} {\bibfnamefont {P.}~\bibnamefont
  {Ghosh}},\ }\bibfield  {title} {\bibinfo {title} {Growth-mediated
  autochemotactic pattern formation in self-propelling bacteria},\ }\href
  {https://doi.org/10.1103/physreve.97.012413} {\bibfield  {journal} {\bibinfo
  {journal} {Phys. Rev. E}\ }\textbf {\bibinfo {volume} {97}},\ \bibinfo
  {pages} {012413} (\bibinfo {year} {2018})}\BibitemShut {NoStop}%
\end{thebibliography}
